\documentclass{elsarticle}
\usepackage[margin=2.5cm]{geometry}

\usepackage[T1]{fontenc}
\usepackage{lmodern}
\usepackage{microtype,url}
\usepackage[dvipsnames]{xcolor}
\usepackage{amsfonts,amsmath,amssymb,bm,mathtools}
\DeclareSymbolFont{vectors}{OML}{lmm}{b}{it}
\DeclareSymbolFont{tensors}{OT1}{lmss}{bx}{it}
\DeclareSymbolFontAlphabet{\tens}{tensors}
\DeclareSymbolFontAlphabet{\mathvec}{vectors} % Vectors
\usepackage[colorlinks=true,allcolors=blue]{hyperref}
\usepackage{siunitx}
\usepackage[noabbrev]{cleveref}

\crefname{section}{}{}
\crefname{equation}{}{}
\crefname{figure}{}{}
\crefname{table}{}{}
\crefname{appendix}{}{}
\crefname{chapter}{}{}
\numberwithin{equation}{section}
\numberwithin{table}{section}
\usepackage[section]{placeins}
\usepackage[doublespacing]{setspace}

\usepackage{graphicx,subcaption}
\usepackage{tikz,pgfplots}
\pgfplotsset{compat=1.16}
\usepgfplotslibrary{statistics}
\usepackage{booktabs}

\DeclareRobustCommand{\d}{\relax\ifmmode\mathrm{d}\else\expandafter\@@d\fi}
\DeclareRobustCommand{\D}{\relax\ifmmode\mathrm{D}\else\expandafter\@@d\fi}
\DeclareRobustCommand{\e}{\relax\ifmmode\mathrm{e}\else\error\fi}
\DeclareRobustCommand{\i}{\relax\ifmmode\mathrm{i}\else\expandafter\@@i\fi}
\DeclareRobustCommand{\uppartial}{\text{\rotatebox[origin=t]{20}{\scalebox{0.95}[1]{\(\partial\)}}}\hspace{-1pt}}

\newcommand{\fd}[2]{\frac{\d #1}{\d #2}}

\newcommand{\fdn}[3]{\frac{\d^#3 #1}{\d #2^#3}}
\newcommand{\pd}[2]{\frac{\uppartial #1}{\uppartial #2}}
\newcommand{\pdt}[3]{\frac{\uppartial^2 #1}{\uppartial #2\uppartial #3}}
\newcommand{\pdn}[3]{\frac{\uppartial^#3 #1}{\uppartial #2^#3}}

\DeclarePairedDelimiterX{\abs}[1]{\lvert}{\rvert}{#1}
\DeclarePairedDelimiterX{\norm}[1]{\lVert}{\rVert}{#1}

\biboptions{sort&compress}

\journal{Physica D: Nonlinear Phenomena.}
\begin{document}
%
%%%%%%%%%%%%%%%%%%%%%%%%
%%%%% Front Matter %%%%%
%%%%%%%%%%%%%%%%%%%%%%%%
%
\begin{frontmatter}
\title{Pattern Formation and Front Stability for a Moving-Boundary Model of Biological Invasion and Recession}
\author[unisa,qut]{Alexander K. Y. Tam}
\author[qut]{Matthew J. Simpson}
\address[unisa]{UniSA STEM, The University of South Australia, Mawson Lakes SA 5095, Australia.}
\address[qut]{School of Mathematical Sciences, Queensland University of Technology, Brisbane QLD 4000, Australia.}
\begin{abstract}
We investigate pattern formation in a two-dimensional (2D) Fisher--Stefan model, which involves solving the Fisher--KPP equation on a compactly-supported region with a moving boundary. The boundary evolves analogously to the classical Stefan problem, such that boundary speed is proportional to the local population density gradient. By combining the Fisher--KPP and classical Stefan theory, the Fisher--Stefan model alleviates two limitations of the Fisher--KPP equation for biological populations. Unlike the Fisher--KPP equation, solutions to the Fisher--Stefan model have compact support, explicitly defining the region occupied by the population. Furthermore, the Fisher--Stefan model admits travelling wave solutions with non-negative density for all wave speeds \(c \in (-\infty,\infty),\) and can thus model both population invasion and population recession. In this work, we investigate whether the 2D Fisher--Stefan model predicts pattern formation, by analysing the linear stability of planar travelling wave solutions to sinusoidal transverse perturbations. Planar fronts of the Fisher--KPP equation are linearly stable. Similarly, we demonstrate that invading planar fronts (\(c > 0\)) of the Fisher--Stefan model are linearly stable to perturbations of all wave numbers. However, our analysis demonstrates that receding planar fronts (\(c < 0\)) of the Fisher--Stefan model are linearly unstable for all wave numbers. This is analogous to unstable solutions for planar solidification in the classical Stefan problem. Introducing a surface tension regularisation stabilises receding fronts for short-wavelength perturbations, giving rise to a range of unstable modes and a most unstable wave number. We supplement linear stability analysis with level-set numerical solutions that corroborate theoretical results. Overall, front instability in the Fisher--Stefan model suggests a new mechanism for pattern formation in receding biological populations.
\end{abstract}
\end{frontmatter}
%
% \linenumbers
%
%%%%%%%%%%%%%%%%%%%%%%%%
%%%%% INTRODUCTION %%%%%
%%%%%%%%%%%%%%%%%%%%%%%%
%
\clearpage
\section{Introduction and Background}\label{sec:intro}
Pattern formation refers to the seemingly spontaneous emergence of visible orderly structures in nature. The science of pattern formation has a long history in applied mathematics, in both biological and industrial applications. In biology, Alan Turing famously introduced the idea that diffusion-driven instability might explain pattern formation in biology~\cite{Turing1952}. His theory showed how reaction and diffusion of chemicals can give rise to stripe, spot, and spiral patterns. Turing's seminal work inspired future research in understanding how reaction--diffusion equations generate pattern formation. In industrial mathematics, one application of pattern formation is melting or freezing at a moving solid--liquid interface. Mullins and Sekerka~\cite{Mullins1964} showed that unidirectional freezing of a dilute binary alloy generates spatial patterns. In this work, we investigate two-dimensional pattern formation in a biological population model with a moving boundary.

The dimensionless Fisher--KPP equation,
\begin{equation}
  \label{eq:intro_fkpp}%
  \pd{u}{t} = \pdn{u}{x}{2} + u\left(1-u\right),
\end{equation}
is a prototype reaction--diffusion model in mathematical biology. The Fisher--KPP model describes a population with density \(u(x,t),\) that disperses via linear Fickian diffusion and proliferates according to a logistic source term~\cite{Kolmogorov1937}. Reaction--diffusion equations often admit travelling-wave solutions on \(-\infty < x < \infty,\) such that the population advances with constant speed and shape. Travelling wave solutions to the Fisher--KPP equation~\cref{eq:intro_fkpp} exist for appropriate boundary conditions, and have long-time speed \(c = 2\) when solved on an infinite domain with compactly-supported initial conditions. Relating properties of the travelling front with population invasion allows the Fisher--KPP equation~\cref{eq:intro_fkpp} and similar reaction--diffusion equations to model phenomena including collective biological cell behaviour~\cite{Gatenby1996,Johnston2015,Maini2004,Sengers2007,Sherratt1990,Simpson2013,Treloar2014}, species invasion in ecology~\cite{Bradshaw-Hajek2004,Broadbridge2022,Skellam1951}, and chemical reactions~\cite{Mercer1995}. However, the Fisher--KPP equation presents two practical challenges when applied to front propagation processes in biology. Firstly, solutions to the Fisher--KPP equation~\cref{eq:intro_fkpp} do not have compact support. Since \(u(x,t) \to 0\) only as \(x \to \infty,\) it is difficult to define the location of the interface between occupied and unoccupied regions of space unambiguously. Secondly, solutions to the Fisher--KPP equation on \(-\infty < x < \infty\) for any initial condition imply population growth and complete colonisation. The Fisher--KPP equation is thus unsuitable for biological phenomena that involve arrested growth~\cite{Landman2003} or recession~\cite{El-Hachem2021}. 

The classical Stefan problem models heat conduction in a material undergoing change of phase, for example ice melting to liquid water~\cite{Rubinstein1971,Gupta2017}. Mathematically, the dimensionless classical Stefan problem is
\begin{subequations}
  \label{eq:intro_stefan}%
  \begin{gather}
    \pd{u}{t} = \pdn{u}{x}{2}, \quad\text{ on }\quad 0 < x < S(t),\label{eq:intro_stefan_heat}\\
    u(0,t) = 1, \quad u(S(t),t) = 0,\label{eq:intro_stefan_bc}\\
    \beta\fd{S}{t} = -\pd{u(S(t),t)}{x},\label{eq:intro_stefan_stefan}\\
    u(x,0) = U(x), \quad\text{ on }\quad x > 0,
  \end{gather}
\end{subequations}
where \(u(x,t)\) is the temperature, \(U(x)\) is the initial temperature distribution, \(S(t)\) is the melted depth, and \(\beta\) is the Stefan number, which is the ratio of latent heat to specific sensible heat. The boundary conditions~\cref{eq:intro_stefan_bc} represent the constant temperature \(u(0,t)\) and the melting temperature \(u(S(t),t).\) These temperatures are typically scaled to be one and zero, respectively. A key feature of the classical Stefan problem is that the interface position between the two phases changes over time according to the one-phase Stefan condition~\cref{eq:intro_stefan_stefan}. The classical Stefan problem is thus a \emph{moving-boundary problem}~\cite{Crank1987}. In biological and industrial mathematics, moving-boundary problems describe phenomena including binary alloy solidification~\cite{Mullins1964,BrosaPlanella2019}, epithelial tissue~\cite{Murphy2021}, cancers~\cite{Ward1997,Jin2021,Shuttleworth2019}, and wound healing~\cite{Zanca2022}.

The Fisher--Stefan model~\cite{Du2010,Du2011,El-Hachem2022} combines the theory of reaction--diffusion equations and moving-boundary problems. This model is similar to the classical Stefan problem~\cref{eq:intro_stefan}, but involves solving the Fisher--KPP equation~\cref{eq:intro_fkpp} instead of the heat equation on the compactly-supported region \(0 < x < L(t).\) We assume that \(L(t)\) evolves according to a Stefan-like condition. The one-dimensional (1D) Fisher--Stefan model~\cite{El-Hachem2022} is
\begin{subequations}
  \label{eq:intro_fs_1d}%
  \begin{gather}
    \pd{u}{t} = \pdn{u}{x}{2} + u\left(1-u\right) \quad\text{ on }\quad 0 < x < L(t),\label{eq:intro_fs_1d_fkpp}\\
    u(0,t) = 1,\quad u(L(t),t) = u_{\mathrm{f}},\label{eq:intro_fs_1d_bc}\\
    \fd{L}{t} = -\kappa\pd{u(L(t),t)}{x},\label{eq:intro_fs_1d_stefan}\\
    u(x,0) = U(x) \quad\text{ on }\quad 0 < x < L(0).\label{eq:intro_fs_1d_ic}
  \end{gather}
\end{subequations}
We use the convention that \(S(t)\) denotes the boundary position in the Stefan problem, and \(L(t)\) the boundary in the Fisher--Stefan model. The parameter \(\kappa\) in~\cref{eq:intro_fs_1d_stefan} relates the gradient of the population density profile at \(L(t)\) to the boundary speed, and is analogous to the inverse Stefan number in the classical Stefan problem~\cref{eq:intro_stefan_stefan}. In the Stefan problem~\cref{eq:intro_stefan}, the temperature at \(x = 0\) is arbitrarily scaled to unity without particular physical meaning. Conversely, the density \(u(0,t) = 1\) in the Fisher--Stefan model does have physical meaning. This density is the value imposed by scaling the carrying capacity density in the logistic source term~\cref{eq:intro_fs_1d_fkpp}. The condition~\cref{eq:intro_fs_1d_bc} \(u(L(t),t) = u_{\mathrm{f}}\) describes the density of a background population with density \(u_{\mathrm{f}},\) which is a constant satisfying \(0 \leq u_{\mathrm{f}} < 1.\) If \(u_{\mathrm{f}} = 0,\) the region \(x \geq L(t)\) is unoccupied, and \(x = L(t)\) defines the interface between occupied and unoccupied regions. Biologically, \(x = L(t)\) might represent the position of a cell invasion front, or the interface between a tumour and healthy tissue. If \(u_{\mathrm{f}} \neq 0,\) we assume there is an inactive background population of constant density that does not diffuse or proliferate. One application of \(u_{\mathrm{f}} \neq 0\) might be a scratch assay experiment in which the region ahead of the main front is not completely vacant. Another interpretation might be fibroblast cells invading a partial wound containing an existing lower-density fibroblast population. The boundary \(x = L(t)\) then represents the interface between members of the population that are active (diffuse and proliferate) and those that are inactive (do not diffuse or proliferate). See El-Hachem, McCue, and Simpson~\cite{El-Hachem2022} for further discussion of the density condition at \(L(t).\)

\begin{figure}
  \centering
  \includegraphics[width=\linewidth]{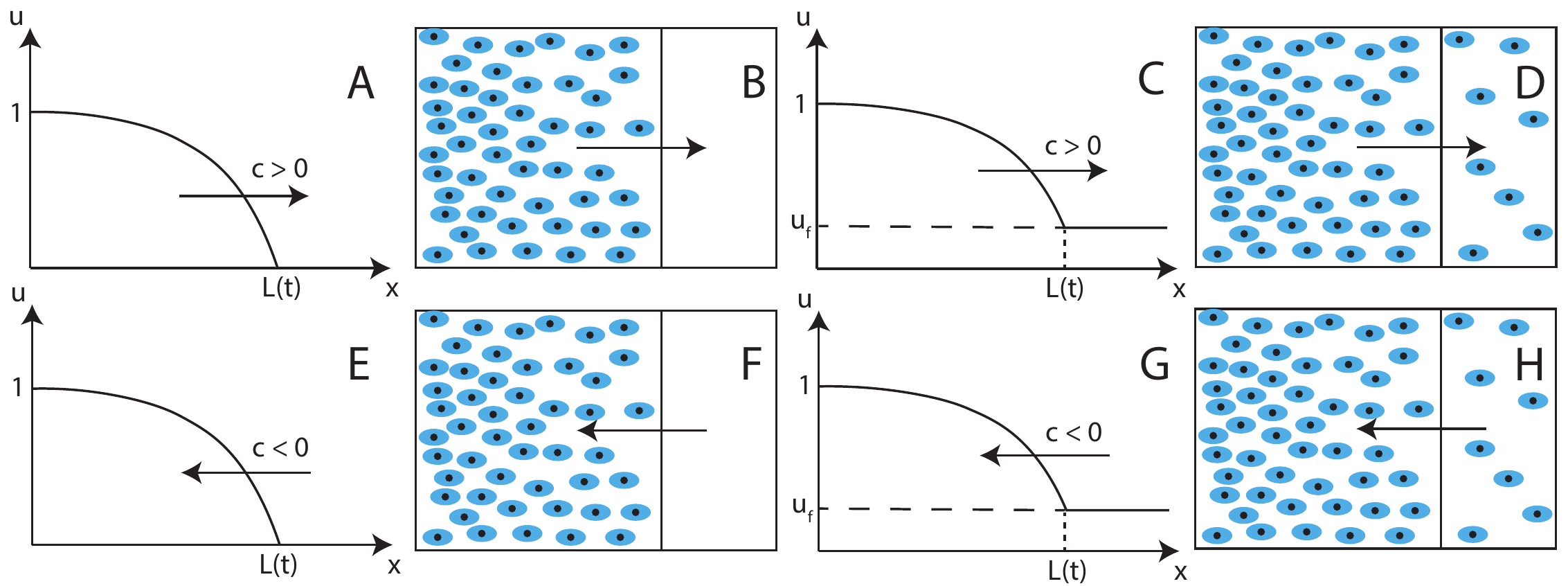}
  \caption{(A--D): Advancing fronts in the 1D Fisher--Stefan model. (A) Advancing front with \(c > 0\) (corresponding to \(\kappa > 0\)) and \(u_{\mathrm{f}} = 0.\) (B) Biological interpretation of an advancing front with \(u_{\mathrm{f}} = 0,\) for example a cell population invading a vacant region. (C) Advancing front with \(c > 0\) (corresponding to \(\kappa > 0\)) and \(0 < u_{\mathrm{f}} \leq 1.\) (D) Biological interpretation of an advancing front with \(u_{\mathrm{f}} \neq 0,\) for example a cell population invading a partially-vacant region. (E--H): Receding fronts in the 1D Fisher--Stefan model. (E) Receding front with \(c < 0\) (corresponding to \(\kappa < 0\)) and \(u_{\mathrm{f}} = 0.\) (F) Biological interpretation of a receding front with \(u_{\mathrm{f}} = 0,\) for example a shrinking tumour. (G) Receding front with \(c < 0\) (corresponding to \(\kappa < 0\)) and \(0 < u_{\mathrm{f}} \leq 1.\) (H) Biological interpretation of a receding front with \(u_{\mathrm{f}} \neq 0.\)}% for example a shrinking tumour leaving behind a residual population.}
  \label{fig:intro_travelling_waves}
\end{figure}

The Fisher--Stefan model~\cref{eq:intro_fs_1d} alleviates the two practical disadvantages of applying the Fisher--KPP equation~\cref{eq:intro_fkpp} to biological populations. The moving boundary \(L(t)\) defines the interface between occupied and unoccupied regions explicitly, and the Fisher--Stefan model~\cref{eq:intro_fs_1d} admits solutions whereby \(L(t)\) decreases, and the population recedes~\cite{El-Hachem2019}. Furthermore, El-Hachem, McCue, and Simpson~\cite{El-Hachem2022} showed that both invading and receding populations can form travelling wave solutions. Their analysis involved introducing a new variable \(z = x - L(t) = x - ct,\) and expressing population density as \(u(z).\) The variable \(z\) follows the moving boundary. For a travelling wave solution, the boundary advances at the constant wave speed \(c,\) such that \(\d L/\d t = c.\) Under this change of variables, the Fisher--Stefan model~\cref{eq:intro_fs_1d} reduces to
\begin{subequations}
  \label{eq:intro_fs_tw}%
  \begin{gather}
    \fdn{u}{z}{2} + c\fd{u}{z} + u(1-u) = 0 \quad\text{ on }\quad -\infty < z < 0,\label{eq:intro_fs_tw_fkpp}\\
    u(-\infty) = 1, \quad u(0) = u_{\mathrm{f}},\label{eq:intro_fs_tw_bc}\\
    -\kappa\fd{u(0)}{z} = c.\label{eq:intro_fs_tw_stefan}
  \end{gather}
\end{subequations}
El-Hachem, McCue, and Simpson~\cite{El-Hachem2022} then used phase plane analysis of~\cref{eq:intro_fs_tw} to show that the Fisher--Stefan model admits travelling wave solutions with non-negative population density for \(-\infty < c < \infty.\) Biologically-feasible (\(u \geq 0\)) receding travelling waves for \(c < 0\) are a distinguishing feature of the Fisher--Stefan model. Most other single-species reaction--diffusion models permit invasion, but do not permit recession. For example, the Fisher--KPP equation~\cref{eq:intro_fkpp} requires \(c \geq 2\) for travelling-wave solutions with non-negative population density~\cite{Canosa1973,Murray2002}. Schematics and biological interpretations of advancing travelling waves are illustrated in Figure~\cref{fig:intro_travelling_waves}A--D, and similar for receding travelling waves are shown in Figure~\cref{fig:intro_travelling_waves}E--H.

An established technique in reaction--diffusion systems and moving-boundary problems is to investigate the linear stability of planar travelling wave or similarity solutions to shape perturbations~\cite{Muller2002,Oelker2017,Mullins1964,Chadam1983,Alert2022}. This analysis extends one-dimensional travelling wave theory to understand pattern formation in two or more spatial dimensions. Linear stability analysis techniques have been applied to patterns in fluid flows~\cite{Mayo2013,Waters2005}, microbial colonies~\cite{Tam2018,Kessler1998,Kitsunezaki1997,Trinschek2018}, chemical reactions~\cite{Merkin2005,Horvath1993,Yang2002,Sivashinsky1977}, and Stefan-type moving-boundary problems~\cite{Langer1980,Rubinstein1982,Strain1988,Doole1996}. M{\"u}ller and van Saarloos~\cite{Muller2002} developed a systematic method to investigate transverse perturbations to a sharp planar interface. The growth rate \(\omega(q)\) of sinusoidal perturbations of wave number \(q\) determines the linear stability (\(\Re(\omega) < 0,\) \emph{i.e.} small-amplitude perturbations decay) or instability (\(\Re(\omega) > 0,\) \emph{i.e.} small-amplitude perturbations grow) of the planar front. The interface considered by M{\"u}ller and van Saarloos~\cite{Muller2002} arose in a coupled reaction--diffusion system with degenerate nonlinear diffusion, but the interface perturbation technique can also apply to reaction--diffusion moving-boundary problems. Notably, planar fronts of the Fisher--KPP equation are linearly stable in multiple spatial dimensions~\cite{Huang2008,Zeng2014}. Using similar analysis, Mullins and Sekerka~\cite{Mullins1964} showed that a planar front is unstable in the classical Stefan problem for planar solidification. Chadam and Ortoleva~\cite{Chadam1983} extended this work by showing that adding surface tension to the interface can stabilise unstable planar solidification fronts.

In this work, we explore pattern formation in the two-dimensional analogue of the Fisher--Stefan model~\cref{eq:intro_fs_1d}, linking previous studies on pattern formation in biological and industrial applications. We use a combination of linear stability analysis and numerical methods. In Section~\cref{sec:model}, we introduce the two-dimensional (2D) Fisher--Stefan model, and present the model in a level-set formulation that underpins our analysis and numerics. In Section~\cref{sec:results}, we present linear stability analysis and numerical results for planar travelling wave fronts subject to transverse sinusoidal perturbations. We show that advancing travelling wave solutions are linearly stable to perturbations of any wave number. This accords with previous studies on the Fisher--KPP equation and the Stefan problem. However, receding travelling wave solutions are linearly unstable for all wave numbers. Like the results of Chadam and Ortoleva~\cite{Chadam1983} for the Stefan problem, applying a surface tension regularisation stabilises receding Fisher--Stefan waves. This enables identification of a most unstable wave number, and characteristic wavelength of pattern formation. Level-set numerical solutions confirm theoretical predictions. Furthermore, numerical solutions extend linear stability theory beyond the small-time regime. Our work provides a new biological interpretation of pattern formation in receding populations.
%
%%%%%%%%%%%%%%%%%%%%%%%%%%%%%%
%%%%% MATHEMATICAL MODEL %%%%%
%%%%%%%%%%%%%%%%%%%%%%%%%%%%%%
%
\section{Two-Dimensional Fisher--Stefan Model}\label{sec:model}
We consider an extension of the Fisher--Stefan model~\cref{eq:intro_fs_1d} to general 2D geometry. Let the region on which we solve the Fisher--KPP equation be \(x < L(y,t),\) such that \(x = L(y,t)\) denotes the interface between the occupied and (partially) vacant regions. An example \(L(y,t)\) for a perturbed planar front is shown in Figure~\cref{fig:model_level-set}A. The 2D Fisher--Stefan model is then
\begin{subequations}
  \label{eq:model_fs_2d}%
  \begin{gather}
    \pd{u}{t} = \pdn{u}{x}{2} + \pdn{u}{y}{2} + u\left(1-u\right), \quad\text{ on }\quad 0 < x < L(y,t),\label{eq:model_fs_2d_fkpp}\\
    u = 1, \quad\text{ on }\quad x = 0\label{eq:model_fs_2d_bc_left},\\
    u = u_{\mathrm{f}}, \quad\text{ on }\quad x = L(y,t)\label{eq:model_fs_2d_bc},\\
    V = -\kappa\nabla u\cdot\mathvec{\hat{n}}, \quad\text{ on }\quad x = L(y,t)\label{eq:model_fs_2d_stefan},\\
    u(x,y,0) = U(x,y), \quad\text{ on }\quad x < L(y,0)\label{eq:model_fs_2d_ic},
  \end{gather}
\end{subequations}
where \(V\) is the speed of the interface, \(\nabla u = (\uppartial_x u, \uppartial_y u)\) is the population density gradient, and \(\mathvec{\hat{n}}\) is the unit outward normal vector to the boundary \(x = L(y,t).\) In practice, we solve~\cref{eq:model_fs_2d} on a finite domain \(0 < y < Y,\) with periodic boundary conditions at \(y = 0\) and \(y = Y.\) %Like the 1D Fisher--Stefan model~\cref{eq:intro_fs_1d}~\cite{Du2010,El-Hachem2019}, the 2D Fisher--Stefan model~\cref{eq:model_fs_2d} admits solutions for population survival and extinction~\cite{Tam2022a}. The survival and extinction behaviour depends on the size and shape of the initial population~\cite{Tam2022a}. 
Before undertaking analysis and numerical solutions, we rewrite the 2D Fisher--Stefan model~\cref{eq:model_fs_2d} in level-set form.
%
%%%%%%%%%%%%%%%%%%%%%%%%%%%%%%%%%
%%%%% Level-Set Formulation %%%%%
%%%%%%%%%%%%%%%%%%%%%%%%%%%%%%%%%
%
\subsection{Level-Set Formulation}\label{sec:model_level-set}
\begin{figure}[htbp!]
  \centering
  \includegraphics[width=\linewidth]{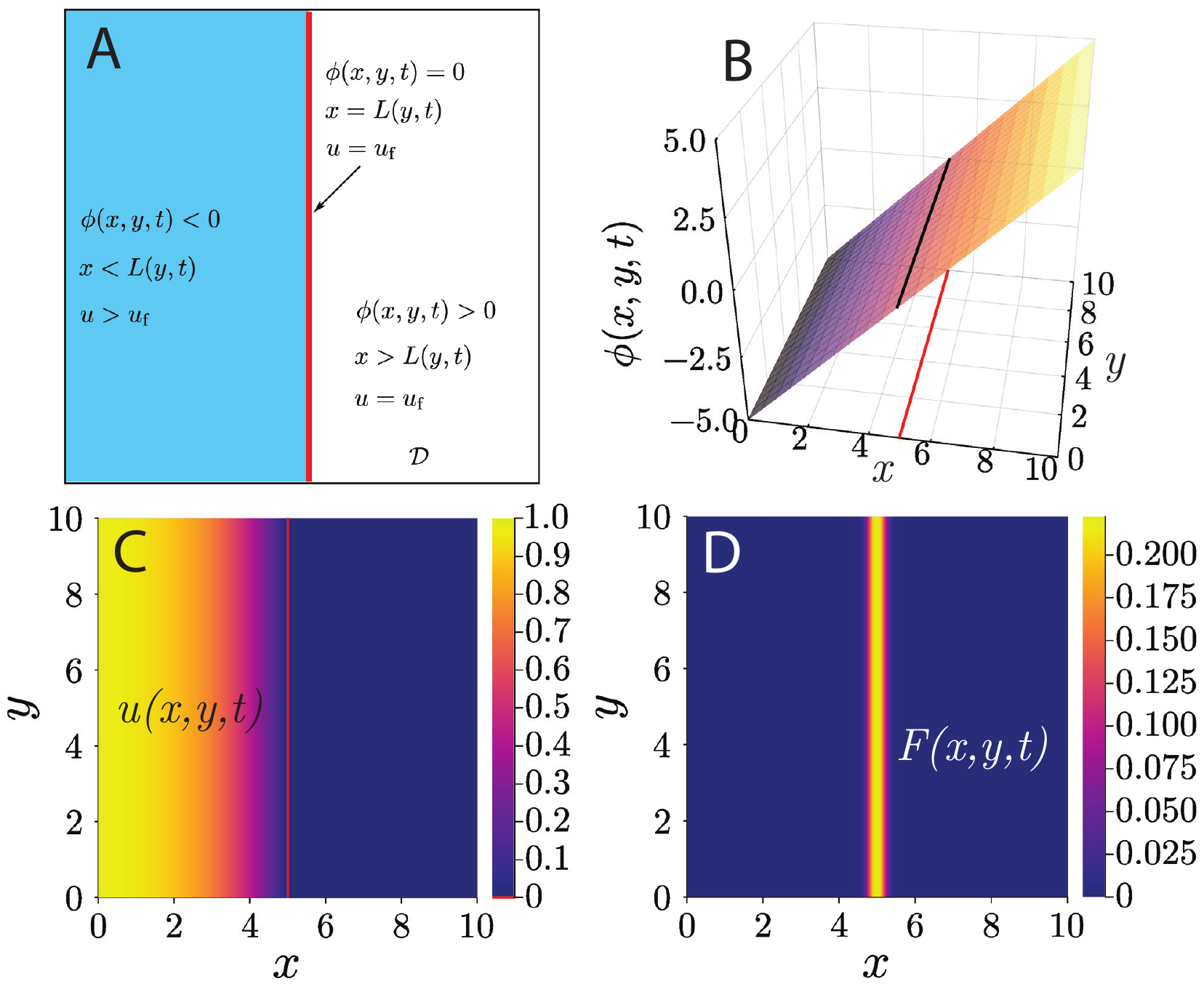}
  \caption{The level-set method for a planar front. (A) Schematic of a computational domain \(\mathcal{D}\) for a planar front. We solve the Fisher--KPP equation on the blue shaded region, \(x < L(y,t).\) The red curve is the boundary, \(x = L(y,t),\) between occupied (blue shaded) and vacant or partially-vacant (non-shaded) regions. (B) A level-set function, \(\phi(x,y,t),\) for a planar front. The black curve is the zero level set, and the red curve is a projection of the zero level set onto the \((x,y)\) plane marked with the grid. (C) A 2D population density profile \(u(x,y,t)\) for a planar front. Red curve denotes the zero level set (interface between occupied and vacant or partially-vacant regions). (D) An extension velocity field \(F(x,y,t)\) for a planar front.}
  \label{fig:model_level-set}
\end{figure}
The level-set form~\cite{Sethian1999,Osher2003} involves embedding the interface position \(x = L(y,t)\) within a new scalar function \(\phi(x,y,t),\) defined such that \(\phi(x,y,t) = 0\) wherever \(x = L(y,t).\) We also impose \(\phi < 0\) wherever \(x < L(y,t),\) and \(\phi \geq 0\) wherever \(x \geq L(y,t).\) A level-set function for a planar travelling wave front is shown in Figure~\cref{fig:model_level-set}B. To maintain \(\phi = 0\) on the interface as it evolves, \(\phi(x,y,t)\) satisfies the level-set equation
\begin{equation}
  \label{eq:model_level-set_equation}%
  \pd{\phi}{t} + F\left\lvert\nabla\phi\right\rvert = 0,
\end{equation}
where \(F(x,y,t)\) is the \emph{extension velocity field}, a scalar function that satisfies \(F = V\) on \(x = L(y,t).\) One possible extension velocity field for a planar travelling wave is shown in Figure~\cref{fig:model_level-set}D. We obtain \(F\) by first approximating \(V\) on the interface using a second-order finite difference approximation, and then applying orthogonal extrapolation as described by Osher and Fedkiw~\cite{Osher2003}. Noting that the unit outward normal is \(\mathvec{\hat{n}} = \nabla\phi/|\nabla\phi|,\) the system
\begin{subequations}
  \label{eq:model_fs_2d_level-set_numerics}%
  \begin{gather}
    \pd{u}{t} = \pdn{u}{x}{2} + \pdn{u}{y}{2} + u\left(1-u\right) \quad\text{ on }\quad \phi(x,y,t) < 0,\label{eq:model_fs_2d_level-set_numerics_fkpp}\\
    \pd{\phi}{t} + F\left\lvert\nabla\phi\right\rvert = 0,\label{eq:model_fs_2d_level-set_numerics_ls}\\
    u = 1 \quad\text{ on }\quad x = 0 \label{eq:model_fs_2d_level-set_numerics_bc_left},\\
    u = u_{\mathrm{f}} \quad\text{ on }\quad \phi(x,y,t) = 0\label{eq:model_fs_2d_level-set_numerics_bc},\\
    F = -\kappa\nabla u\cdot\frac{\nabla\phi}{\left\lvert\nabla\phi\right\rvert} \quad\text{ on }\quad \phi(x,y,t) = 0\label{eq:model_fs_2d_level-set_numerics_velocity},\\
    u(x,y,0) = U(x,y) \quad\text{ on }\quad \phi(x,y,0)\label{eq:model_fs_2d_level-set_numerics_ic},
  \end{gather}
\end{subequations}
is equivalent to the 2D Fisher--Stefan model~\cref{eq:model_fs_2d}, but avoids the need to find \(L(y,t)\) explicitly. Our numerical level-set method involves solving the system~\cref{eq:model_fs_2d_level-set_numerics} using explicit finite-difference methods, with automatic time-stepping to control truncation error~\cite{Tam2022a,Jiang1998,Aslam2004,Tsitouras2011,Rackauckas2017,Morrow2021}. We always ensure the interface is sufficiently far from the left boundary \(x = 0,\) such that the Dirichlet condition~\cref{eq:model_fs_2d_level-set_numerics_bc_left} approximates the far-field density. We solve~\cref{eq:model_fs_2d_level-set_numerics} on \(0 < y < Y,\) with periodic boundary conditions at \(y = 0\) and \(y = Y.\) Full details of our level-set implementation are provided in~\cref{app:numerics_level-set} (Supplementary Data), and open-source \textsc{Julia} code is freely available on \href{https://github.com/alex-tam/2D_Fisher-Stefan_Level-Set_Stability}{GitHub}. 

As we will see in Section~\cref{sec:results}, the level-set formulation also provides a natural setting for the linear stability analysis. Combining the level-set equation~\cref{eq:model_fs_2d_level-set_numerics_ls} with the condition for the extension velocity field on the interface~\cref{eq:model_fs_2d_level-set_numerics_velocity}, we obtain a problem in similar form to that of Chadam and Ortoleva~\cite{Chadam1983},
\begin{subequations}
  \label{eq:model_fs_2d_level-set_analysis}%
  \begin{gather}
    \pd{u}{t} = \pdn{u}{x}{2} + \pdn{u}{y}{2} + u\left(1-u\right) \quad\text{ on }\quad \phi(x,y,t) < 0,\label{eq:model_fs_2d_level-set_analysis_fkpp}\\
    \pd{\phi}{t} = \kappa\nabla u\cdot\nabla\phi \quad\text{ on }\quad \phi(x,y,t) = 0,\label{eq:model_fs_2d_level-set_analysis_ls}\\
    u = 1 \quad\text{ on }\quad x = 0\label{eq:model_fs_2d_level-set_analysis_bc_left},\\
    u = u_{\mathrm{f}} \quad\text{ on }\quad \phi(x,y,t) = 0\label{eq:model_fs_2d_level-set_analysis_bc},\\
    u(x,y,0) = U(x,y) \quad\text{ on }\quad \phi(x,y,0) < 0.\label{eq:model_fs_2d_level-set_analysis_ic}
  \end{gather}
\end{subequations}
The level-set formulation~\cref{eq:model_fs_2d_level-set_analysis} converts the 2D Fisher--Stefan model~\cref{eq:model_fs_2d} from a problem on \(0 < x < L(y,t)\) to a problem defined for \(-L(y,t) < \phi < 0.\) The system~\cref{eq:model_fs_2d_level-set_analysis} underpins the linear stability analysis presented in Section~\cref{sec:results}.
%
%%%%%%%%%%%%%%%%%%%%%%%%%%%%%%%%%%%%%
%%%%% LINEAR STABILITY ANALYSIS %%%%%
%%%%%%%%%%%%%%%%%%%%%%%%%%%%%%%%%%%%%
%
\section{Linear Stability Analysis and Numerical Results}\label{sec:results}
\begin{figure}[htbp!]
  \centering
  \includegraphics[width=\linewidth]{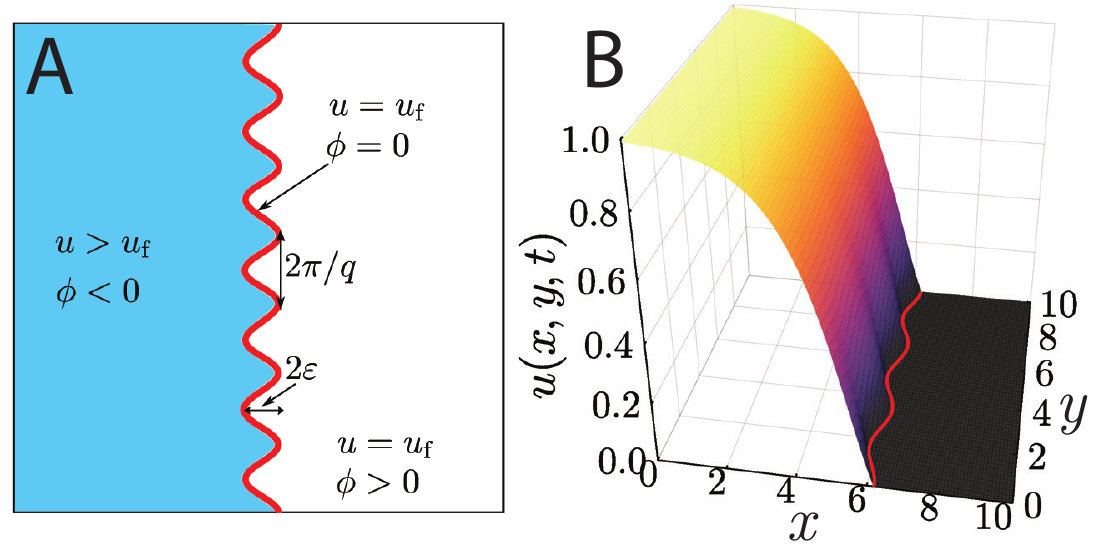}
  \caption{The linear stability ansatz for analysing two-dimensional pattern formation using transverse perturbations. (A) Schematic of a perturbed planar front, illustrating the perturbation amplitude, \(\varepsilon,\) and perturbation wave number, \(q.\) Red curve denotes the zero level set (interface between occupied and vacant or partially-vacant regions). (B) Density profile for a perturbed planar travelling wave with \(\varepsilon = 0.2,\) \(q = \pi,\) \(\beta = 6,\) \(\kappa = 0.5,\) \(u_{\mathrm{f}} = 0.\) Red curve denotes the zero level set.}
  \label{fig:results_ansatz}
\end{figure}

We investigate the linear stability of planar travelling wave solutions to the Fisher--Stefan model. Following M{\"u}ller and van Saarloos~\cite{Muller2002}, we expand the front position as
\begin{equation}
  \label{eq:results_perturbation_front}%
  L(y,t) = ct + \varepsilon\e^{\i qy + \omega t} + \mathcal{O}(\varepsilon^2),
\end{equation}
where \(\varepsilon \ll 1,\) \(q\) is the wave number of the perturbation, and \(\omega\) is the growth rate. The ansatz~\cref{eq:results_perturbation_front} corresponds to small-amplitude sinusoidal perturbations to the interface position of planar travelling waves moving with speed \(c.\) That is, an advancing (right-moving) wave corresponds to \(c > 0,\) and a receding (leftward-moving) wave corresponds to \(c < 0.\) To aid the analysis, we replace the independent variable \(x\) with
\begin{equation}
  \label{eq:results_xi}%
  \xi = x - ct - \varepsilon\e^{\i qy + \omega t},
\end{equation}
such that \(\xi = 0\) follows the perturbed moving boundary up to \(\mathcal{O}(\varepsilon).\) We assume that the far-field density to the left of the front (\(\xi \to -\infty,\)) is \(u = 1\) for both advancing and receding travelling waves, as per the boundary condition~\cref{eq:model_fs_2d_level-set_analysis_bc_left}. The level-set function then satisfies \(\phi = x - ct - \varepsilon\e^{\i qy + \omega t} + \mathcal{O}(\varepsilon^2),\) such that \(\phi = 0\) wherever \(\xi = 0.\) In terms of the independent variables \((\xi,y,t),\) we also expand the population density as
\begin{equation}
  \label{eq:results_perturbation_density}%
  u(\xi,y,t) = u_0(\xi) + \varepsilon u_1(\xi)\e^{\i qy + \omega t} + \mathcal{O}(\varepsilon^2).
\end{equation}
This, and the front position perturbation~\cref{eq:results_perturbation_front}, are illustrated in Figure~\cref{fig:results_ansatz}.

On applying the change of variables and expansions~\cref{eq:results_perturbation_front,eq:results_xi,eq:results_perturbation_density}, the 2D Fisher--Stefan model~\cref{eq:model_fs_2d_level-set_analysis} then yields, to leading-order
\begin{subequations}
  \label{eq:results_lo}%
  \begin{gather}
    \fdn{u_0}{\xi}{2} + c\fd{u_0}{\xi} + u_0\left(1-u_0\right) = 0 \quad\text{ on }\quad \xi < 0,\label{eq:results_lo_fkpp}\\
    u_0(-\infty) = 1, \quad u_0(0) = u_{\mathrm{f}},\label{eq:results_lo_bc}\\
    \fd{u_0(0)}{\xi} = -\frac{c}{\kappa}.\label{eq:results_lo_stefan}
  \end{gather}
\end{subequations}
These are the 1D Fisher--Stefan travelling-wave equations~\cref{eq:intro_fs_tw} for \(u_0.\) Thus, leading-order solutions to the 2D Fisher--Stefan model~\cref{eq:model_fs_2d_level-set_analysis} with the ansatzes~\cref{eq:results_perturbation_front,eq:results_perturbation_density} are planar travelling waves. The system~\cref{eq:results_lo} is a two-point second-order boundary-value problem, with three boundary conditions and an unknown parameter, \(c.\) We solve~\cref{eq:results_lo} by approximating the infinite domain \(-\infty < \xi < 0\) with a truncated but sufficiently large finite domain \(-\xi_{\mathrm{max}} < \xi < 0,\) and use central finite-difference schemes to discretise~\cref{eq:results_lo_fkpp,eq:results_lo_bc} on \(-\xi_{\mathrm{max}} < \xi < 0.\) This yields a system of nonlinear algebraic equations, which we solve using the Newton--Raphson method. We then use the shooting method to obtain the value of \(c\) that satisfies~\cref{eq:results_lo_stefan}. Full details on our numerical methods for boundary-value problems are provided in~\cref{app:numerics_shooting}.

\begin{figure}[htbp!]
  \centering
  \includegraphics[width=0.75\linewidth]{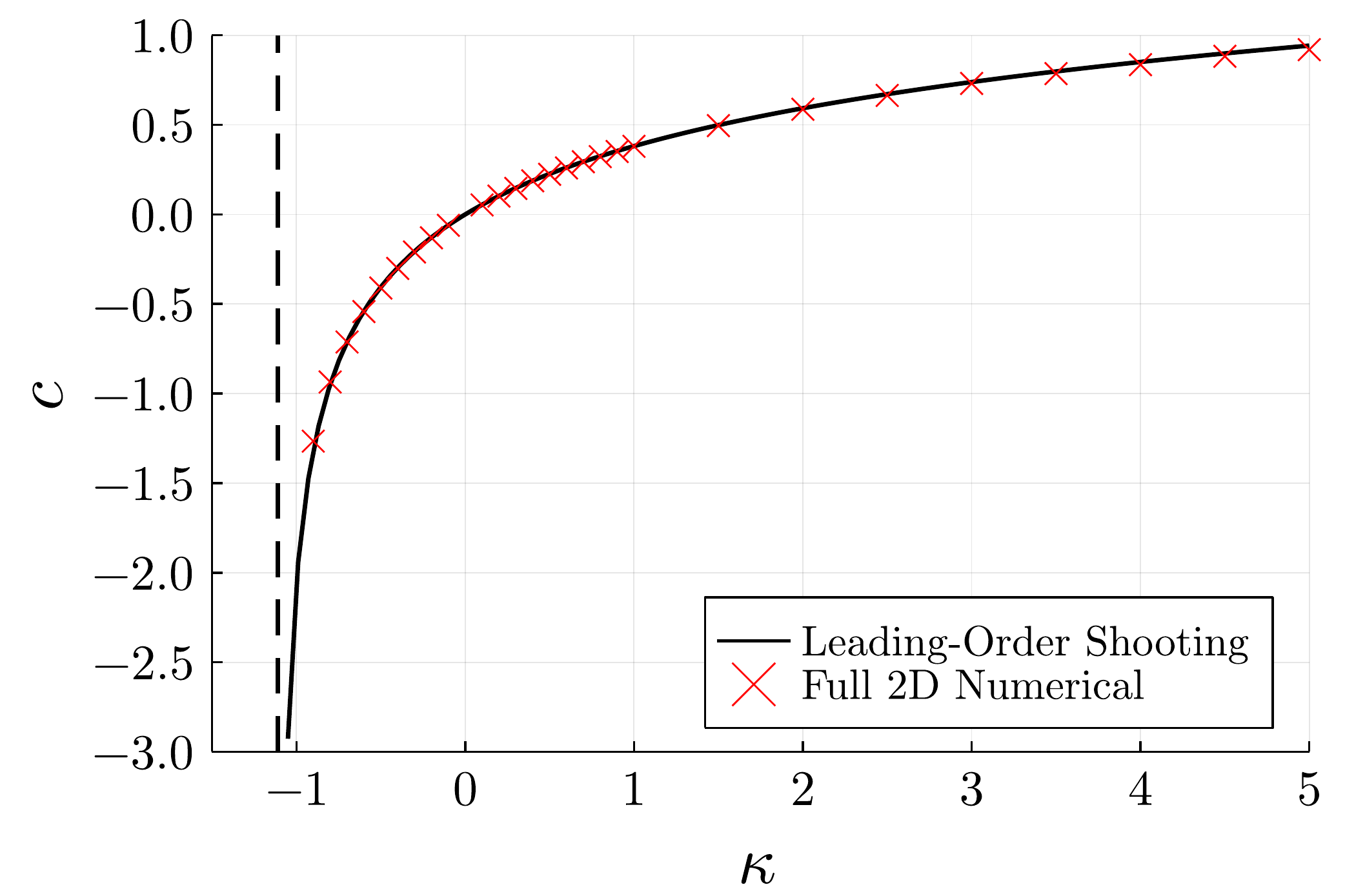}
  \caption{The effect of \(\kappa\) on the travelling wave speed, \(c,\) in the Fisher--Stefan model with \(u_{\mathrm{f}} = 0.1.\) The solid curve represents theoretical results from solving~\cref{eq:results_lo} using the shooting method. Crosses are numerical results obtained from level-set solutions to the 2D Fisher--Stefan model~\cref{eq:model_fs_2d}. Full 2D level-set numerical results were obtained with \(\Delta x = \Delta y = 0.1\) and \(\Delta t = 0.002,\) the solution to~\cref{eq:results_lo} as the initial conditions, and a solution duration of \(T = 1.\) The dashed line is the vertical asymptote \(\kappa = -1/(1-u_{\mathrm{f}}).\)}
  \label{fig:results_wave_speed}
\end{figure}

Our numerical shooting procedure for the leading-order problem~\cref{eq:results_lo} solves for the travelling wave profile, \(u_0(\xi),\) and the wave speed, \(c,\) for given fixed \(\kappa\) and \(u_{\mathrm{f}}.\) The relationship between \(c\) and \(\kappa\) for varying \(u_{\mathrm{f}}\) is shown in Figure~\cref{fig:results_wave_speed}. For any value of \(u_{\mathrm{f}},\) advancing travelling waves (\(c > 0\)) correspond to \(\kappa > 0,\) and receding travelling waves (\(c < 0\)) correspond to \(\kappa < 0.\) Furthermore, for \(0 \leq u_{\mathrm{f}} < 1\) the wave speed blows up \(c \to -\infty\) as \(\kappa \to -1/(1-u_{\mathrm{f}})\) from above. This blow-up was recently demonstrated and analysed by El-Hachem, McCue, and Simpson~\cite{El-Hachem2021,El-Hachem2022}. Hence, \(u_{\mathrm{f}}\) governs the minimum value of \(\kappa\) needed to obtain travelling wave solutions with finite wave speed. The vertical asymptote in Figure~\cref{fig:results_wave_speed} demonstrates this property for \(u_{\mathrm{f}} = 0.1.\)
We compare shooting method results for leading-order problem~\cref{eq:results_lo} with full level-set numerical solutions of the 2D Fisher--Stefan model. In full numerical solutions, we solve the Fisher--Stefan model on \((x,y) \in [0,10] \times [0,10],\) using \(\Delta x = \Delta y = 0.1.\) We use the initial condition \(u(x,y,0) = u_0(x-5),\) where \(u_0\) is the leading-order solution to~\cref{eq:results_lo} obtained using the shooting method. This corresponds to a travelling wave with its sharp front at \(x = 5.\) We compute the solution until \(t = T\) using \(\Delta t = 0.002,\) and obtain the front position \(L(5, T)\) by finding the value of \(x\) such that \(\phi(x,5,T) = 0\) using linear interpolation. We then estimate the wave speed using \(c_{\mathrm{num}} = (L(5,T) - 5)/T,\) for \(T = 1.\) These results are given as crosses in Figure~\cref{fig:results_wave_speed}, and confirm the numerical method can reproduce the analytical wave speed for a range of \(\kappa.\)

The leading-order solution to~\cref{eq:results_lo} determines the base-state that we perturb in our analysis. The \(\mathcal{O}(\varepsilon)\) problem determines the linear stability of planar travelling waves to transverse perturbations. Under the same ansatzes~\cref{eq:results_perturbation_front,eq:results_perturbation_density}, the \(\mathcal{O}(\varepsilon)\) problem for the first-order correction terms \(u_1(\xi)\) is
\begin{subequations}
  \label{eq:results_fo}%
  \begin{gather}
    \fdn{u_1}{\xi}{2} + c\fd{u_1}{\xi} + \left[1-\omega - q^2 - 2u_0(\xi)\right]u_1(\xi) = -\left(\omega + q^2\right)\fd{u_0}{\xi} \quad\text{ on }\quad \xi < 0,\label{eq:results_fo_fkpp}\\
    u_1(-\infty) = 0, \quad u_1(0) = 0,\label{eq:results_fo_bc}\\
    \fd{u_1(0)}{\xi} = -\frac{\omega}{\kappa}.\label{eq:results_fo_stefan}
  \end{gather}
\end{subequations}
Like~\cref{eq:results_lo}, the system~\cref{eq:results_fo} is an overdetermined second-order boundary-value problem with three boundary conditions. To solve~\cref{eq:results_fo}, we fix the wave number \(q\) and apply the shooting method on \(-\xi_{\mathrm{max}} \leq \xi \leq 0\) to solve for \(u_1(\xi)\) and the growth rate \(\omega.\) The shooting method is described fully in~\cref{app:numerics_shooting}. We repeat this numerical process for varying \(q,\) and obtain the dispersion relation \(\omega(q).\) A planar wave is linearly unstable to perturbations with wave number \(q\) if \(\Re(\omega(q)) > 0,\) and linearly stable if \(\Re(\omega(q)) < 0.\) The dispersion relation \(\omega(q)\) provides the range of wave numbers \(q\) for which the wave is unstable. If there is instability, it also provides the \emph{most unstable wave number}, the wave number \(q\) with the largest growth rate. For random perturbations with multiple modes, perturbations of the most unstable wave number will outcompete those of other wave numbers, and eventually dominate the resulting spatial pattern. We use this linear stability analysis to characterise pattern formation for advancing and receding travelling waves, and receding waves with a surface tension regularisation.
%
%%%%%%%%%%%%%%%%%%%%%%%%%%%%%%%%%%%%
%%%%% Results: Advancing Waves %%%%%
%%%%%%%%%%%%%%%%%%%%%%%%%%%%%%%%%%%%
%
\subsection{Advancing Travelling Waves are Linearly Stable}\label{ssec:results_advancing}
\begin{figure}[htbp!]
  \centering
  \includegraphics[width=\linewidth]{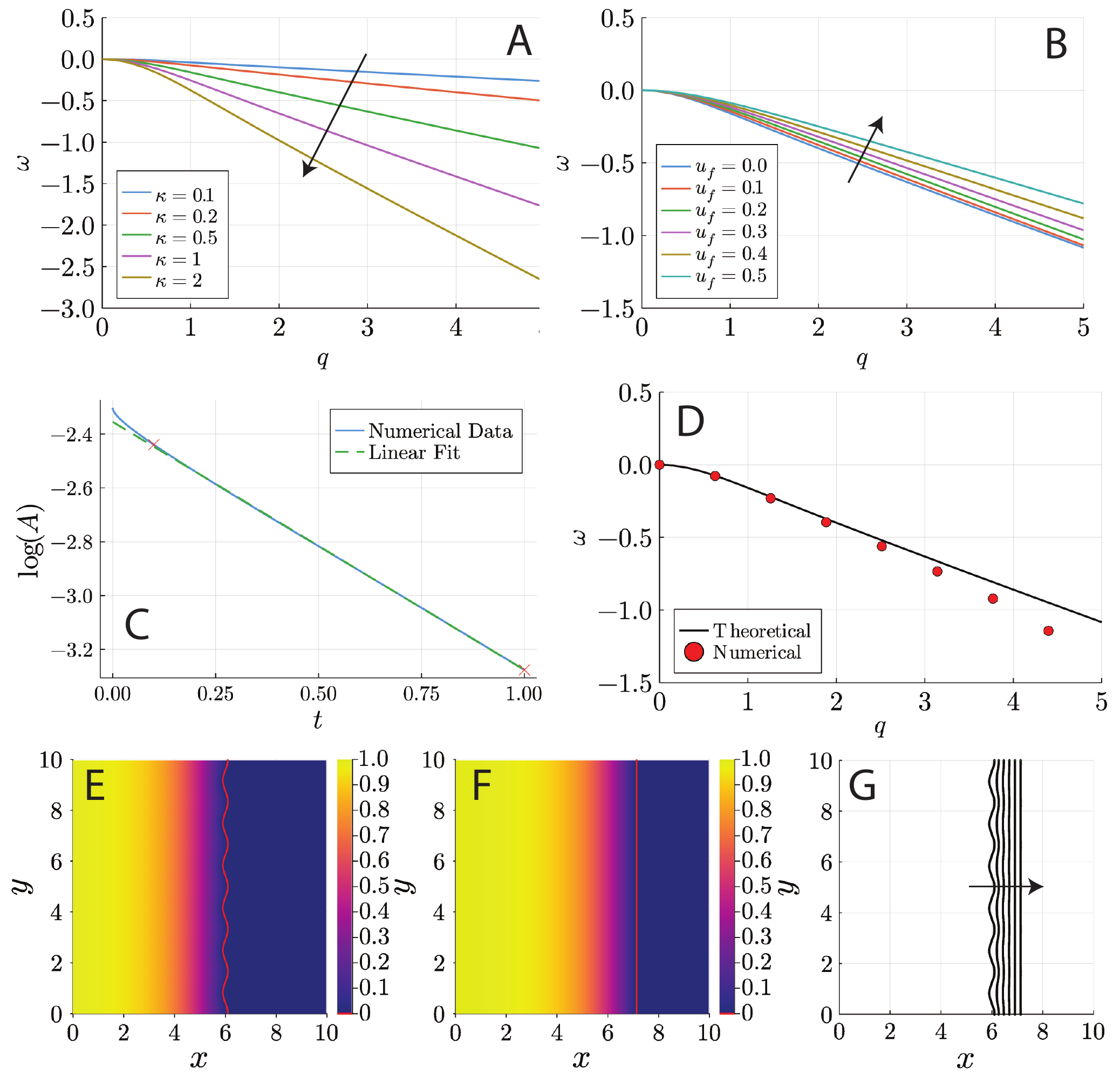}
  \caption{Linear stability analysis results for advancing fronts (\(\kappa > 0\)). (A--B): Dispersion relations shows the relationship between wave number, \(q,\) and growth rate \(\omega.\) For each positive wave number \(q > 0,\) the growth rate \(\omega < 0,\) indicating that the front is linearly stable to transverse perturbations. Arrows indicate direction of increasing parameter. (A) Solutions with varying \(\kappa\) and \(u_{\mathrm{f}} = 0.\) (B) Solutions with \(\kappa = -0.5\) and varying \(u_{\mathrm{f}}.\) (C--D): Comparison between theoretical and numerical dispersion relation. (C) Linear regression to determine estimate the growth rate for \(\kappa = 0.5,\) \(u_{\mathrm{f}} = 0,\) and \(q = 6\pi/5.\) Data between the two crosses are used for the linear fit. (D) Dispersion relation for \(\kappa = 0.5,\) and \(u_{\mathrm{f}} = 0.\) Dots indicate growth rate estimated from 2D level-set numerical solutions, and the solid curve is the theoretical result. (E--G): Numerical solution to the Fisher--Stefan model with \(\kappa = 0.5,\) \(u_{\mathrm{f}} = 0,\) and \(q = 6\pi/5.\) Red curves denote the zero level set (interface between occupied and vacant or partially-vacant regions). (E) Initial condition~\cref{eq:results_numerical_ic}, with \(\varepsilon = 0.1,\) \(q = 6\pi/5,\) and \(\beta = 6.\) (F) Numerical solution \(u(x,y,5).\) (G) Interface plotted for \(t \in \{0,1,2,3,4,5\},\) showing front stability. Arrow indicates increasing time.}
  \label{fig:results_lsa_advancing}
\end{figure}

We first investigate the linear stability of advancing planar waves with \(\kappa > 0\) and \(c > 0.\) Figures~\cref{fig:results_lsa_advancing}A--B illustrate the dispersion relations \(\omega(q).\) In Figure~\cref{fig:results_lsa_advancing}A, we set \(u_{\mathrm{f}} = 0\) and vary \(\kappa,\) and in Figure~\cref{fig:results_lsa_advancing}B, we set \(\kappa = 0.5\) and vary \(u_{\mathrm{f}}.\) For each value of \(\kappa\) and \(u_{\mathrm{f}}\) shown, planar Fisher--Stefan travelling waves are linearly stable to perturbations of all wave numbers. This finding accords with linear stability analysis of the Fisher--KPP equation, for which planar fronts linearly stable~\cite{Huang2008,Zeng2014}. Linear stability for all wave numbers is also consistent with results for melting in the Stefan problem~\cite{Chadam1983}, which corresponds to \(\kappa > 0\) in the Fisher--Stefan model.

Numerical solutions to the full 2D Fisher--Stefan model extend the linear stability analysis beyond the small-time, small-perturbation regime. To confirm the analysis in Figure~\cref{fig:results_lsa_advancing}A--B, we computed solutions using the level-set method, with \(\kappa = 0.5,\) \(u_\mathrm{f} = 0,\) and varying \(q.\) We used the initial condition,
\begin{equation}
  \label{eq:results_numerical_ic}%
  u(x,y,0) = u_0\left[x - \beta - \varepsilon\cos(qy)\right] + \varepsilon u_1\left[x - \beta - \varepsilon\cos(qy)\right]\cos(qy),
\end{equation}
which matches the perturbation ansatz~\cref{eq:results_perturbation_density}, and we use \(\varepsilon = 0.1.\) In~\cref{eq:results_numerical_ic}, \(u_0\) and \(u_1\) are the respective solutions to~\cref{eq:results_lo,eq:results_fo} obtained using the shooting method, and \(\beta\) is a constant that applies a horizontal translation. Throughout this work, we compute numerical solutions with perturbed planar fronts, using \(\Delta x = \Delta y = 0.025,\) and \(\Delta t = 0.001.\) Equation~\cref{eq:results_numerical_ic} represents a planar travelling wave front with transverse sinusoidal perturbations of amplitude \(\varepsilon\) and wave number \(q.\) The perturbation amplitude evolves with time. At each time step in the numerical solution, we measure the perturbation amplitude
\begin{equation}
  \label{eq:results_numerical_perturbation_amplitude}%
  A(t) = \frac{X_{\mathrm{max}}(t) - X_{\mathrm{min}}(t)}{2},
\end{equation}
where
\begin{subequations}
  \label{eq:results_numerical_X}%
  \begin{gather}
    X_{\mathrm{max}}(t) = \max_{y} \{ x \mid \phi(x,y,t) = 0\},\label{eq:results_numerical_X_max}\\
    X_{\mathrm{min}}(t) = \min_{y} \{ x \mid \phi(x,y,t) = 0\}.\label{eq:results_numerical_X_min}
  \end{gather}
\end{subequations}
In practice, we use linear interpolation to find the positions \(x_j\) such that \(\phi(x,y_j,t_k) = 0\) for every \(j = 0, \dots, N_y,\) at given time \(t = t_k.\) The maximum and minimum values in the set \(\{x_j\}\) then determine~\cref{eq:results_numerical_X}, and subsequently the numerical perturbation amplitude~\cref{eq:results_numerical_perturbation_amplitude}. To estimate the growth rate \(\omega,\) we assume the ansatz~\cref{eq:results_perturbation_front} such that the perturbation amplitude~\cref{eq:results_numerical_perturbation_amplitude} grows or decays exponentially. We then use \nolinkurl{Polynomials.jl} to obtain a linear fit \(\log(A) = \omega_{\mathrm{num}}t + C,\) where \(C\) is a constant and \(\omega_{\mathrm{num}}\) is the growth rate estimated from numerical data. In practice, we use data for \(t \in [0.1,1]\) to obtain \(\omega_{\mathrm{num}}.\) As Figure~\cref{fig:results_lsa_advancing}C shows, this range gives an approximately linear relationship between \(\log(A)\) and \(t\) in the numerical solutions. Thus, the assumption of exponential growth or decay in perturbation amplitude remains valid beyond the small-time regime.

We computed numerical solutions with \(\kappa = 0.5,\) \(u_{\mathrm{f}} = 0,\) and various \(q.\) In each solution, the amplitude of the initial sinusoidal perturbations decayed, corroborating the linear stability results. As Figure~\cref{fig:results_lsa_advancing}D shows, the numerical estimate agrees well with the theoretical calculation, particularly for small \(q\) (long-wavelength perturbations). A possible explanation for the discrepancy for larger \(q\) is that a smaller grid spacing is required to accurately resolve short-wavelength perturbations. Since our level-set scheme involves explicit finite-difference approximations, decreasing the grid spacing further would require prohibitively short time steps to maintain numerical stability. Furthermore, our choice of \(\varepsilon = 0.1\) might not be sufficiently small to match linear theory. In our numerical solutions, we cannot choose arbitrarily small \(\varepsilon,\) because the perturbation amplitude must exceed the grid spacing. This might provide another explanation for the discrepancy between theory and numerical experiments at large \(q.\)

Numerical solutions can also explore pattern formation beyond the small-time linear regime. An example numerical solution with \(\kappa = 0.5,\) \(u_{\mathrm{f}} = 0,\) and \(q = 6\pi/5\) is shown in Figures~\cref{fig:results_lsa_advancing}E--F. Figure~\cref{fig:results_lsa_advancing}E shows the perturbed planar front initial condition, and Figure~\cref{fig:results_lsa_advancing}F shows the solution at \(t = 5.\) The initial perturbations decayed, giving rise to a planar front that persists for long time. These results suggest that advancing fronts remain stable to transverse perturbations beyond the linear regime.
%
%%%%%%%%%%%%%%%%%%%%%%%%%%%%%%%%%%%
%%%%% Results: Receding Waves %%%%%
%%%%%%%%%%%%%%%%%%%%%%%%%%%%%%%%%%%
%
\subsection{Receding Travelling Waves are Linearly Unstable}\label{ssec:results_receding}
\begin{figure}[htbp!]
  \centering
  \includegraphics[width=\linewidth]{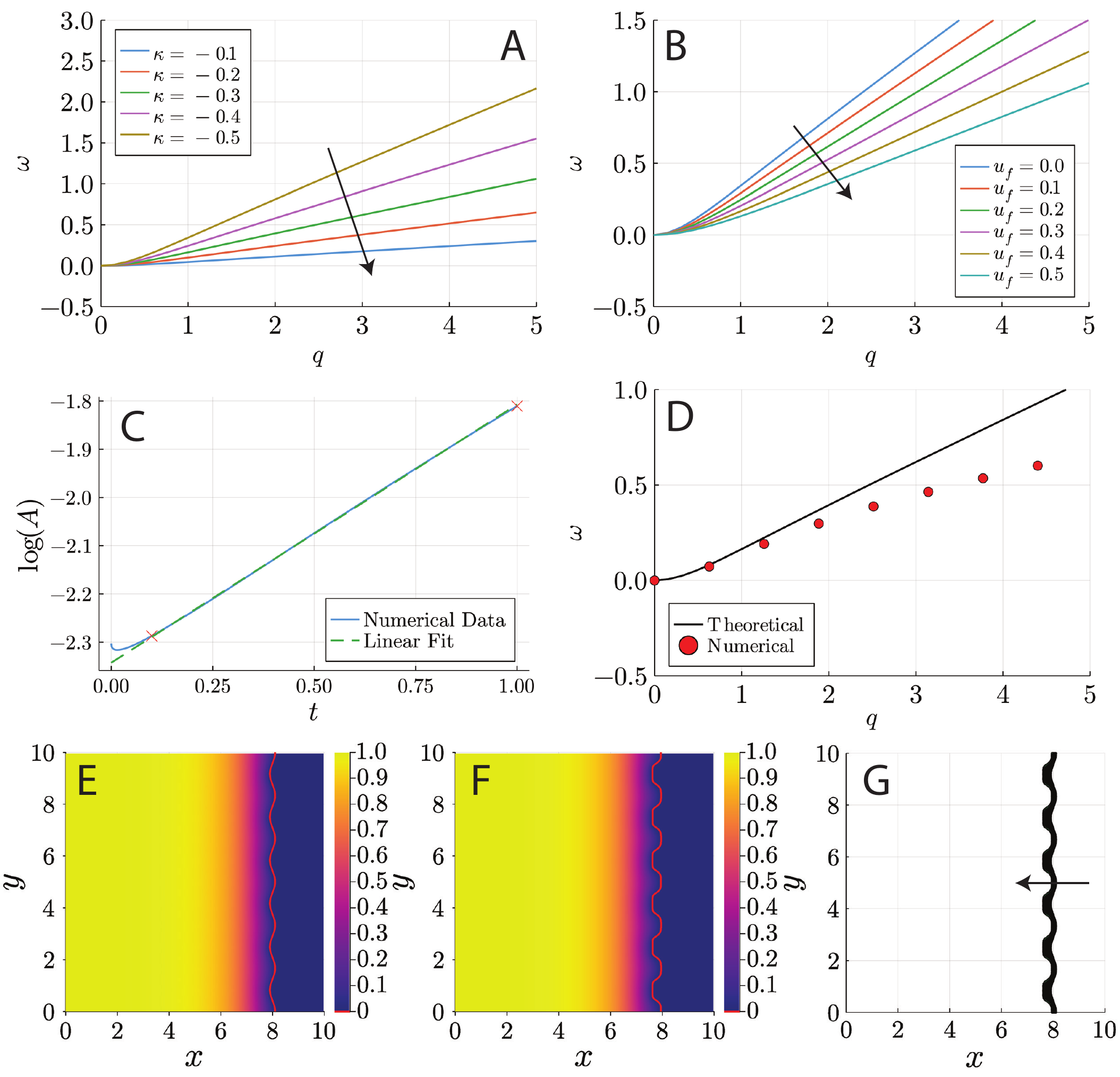}
  \caption{Linear stability analysis results for receding fronts (\(\kappa < 0\)). (A--B): Dispersion relations shows the relationship between wave number, \(q,\) and growth rate \(\omega.\) For each positive wave number \(q > 0,\) the growth rate \(\omega < 0,\) indicating that the front is linearly stable to transverse perturbations. Arrows indicate direction of increasing parameter. (A) Solutions with varying \(\kappa\) and \(u_{\mathrm{f}} = 0.\) (B) Solutions with \(\kappa = -0.3\) and varying \(u_{\mathrm{f}}.\) (C--D): Comparison between theoretical and numerical dispersion relation. (C) Linear regression to determine estimate the growth rate for \(\kappa = -0.3,\) \(u_{\mathrm{f}} = 0,\) and \(q = 6\pi/5.\) Data between the two crosses are used for the linear fit. (D) Dispersion relation for \(\kappa = -0.3,\) and \(u_{\mathrm{f}} = 0.\) Dots indicate growth rate estimated from 2D level-set numerical solutions, and the solid curve is the theoretical result. (E--G): Numerical solution to the Fisher--Stefan model with \(\kappa = -0.3,\) \(u_{\mathrm{f}} = 0,\) and \(q = 6\pi/5.\) Red curves denote the zero level set (interface between occupied and vacant or partially vacant regions). (E) Initial condition~\cref{eq:results_numerical_ic}, with \(\varepsilon = 0.1,\) \(q = 6\pi/5,\) and \(\beta = 8.\) (F) Numerical solution \(u(x,y,1).\) (G) Interface plotted for \(t \in \{0, 0.2, 0.4, 0.6, 0.8, 1\},\) showing front instability. Arrow indicates increasing time.}
  \label{fig:results_lsa_receding}
\end{figure}

In contrast to the advancing fronts investigated in Section~\cref{ssec:results_advancing}, receding planar fronts are linearly unstable to transverse perturbations of all wave numbers. Figures~\cref{fig:results_lsa_receding}A--B show dispersion relations for receding fronts. In Figure~\cref{fig:results_lsa_receding}A, we set \(u_{\mathrm{f}} = 0\) and vary \(\kappa,\) and in Figure~\cref{fig:results_lsa_receding}B we set \(\kappa = -0.5\) and vary \(u_{\mathrm{f}}.\) For all values of \(\kappa\) and \(u_{\mathrm{f}}\) presented, the growth rate \(\omega > 0\) for every wave number \(q > 0,\) indicating linear instability. The function \(\omega(q)\) increases with increasing \(q\) for all combinations of \(\kappa\) and \(u_{\mathrm{f}}\) tested. Our theory thus predicts short-wavelength (large \(q\)) perturbations to grow faster than long-wavelength perturbations. The most unstable perturbations occur as \(q \to \infty,\) such that there is no finite most-unstable wave number.

The possibility of linear instability distinguishes the Fisher--Stefan model from the Fisher--KPP equation. As our work shows, linear instability in the Fisher--Stefan model only occurs for receding fronts with \(\kappa < 0.\) In contrast, planar front solutions to the Fisher--KPP equation~\cite{Huang2008,Zeng2014}, and the Fisher--Stefan model with \(\kappa > 0,\) advance and are linearly stable. Receding fronts in the Fisher--Stefan model are analogous to planar solidification in the Stefan problem. Our finding of linear instability is consistent with previous studies~\cite{Rubinstein1982,Chadam1983} showing that planar solidification is unstable. Linear instability in the Fisher--Stefan model provides a possible explanation for biological pattern formation that is unavailable in the standard Fisher--KPP model.

We computed numerical level-set solutions to validate this linear instability. Figure~\cref{fig:results_lsa_receding}D compares the theoretical stability curve for \(\kappa = -0.3,\) and \(u_{\mathrm{f}} = 0.0.\) Dots indicate estimates based on level-set numerical solutions (using the method described in Section~\cref{ssec:results_advancing}), and the solid curve is the theoretical prediction obtained using the shooting method. Theory and numerical solutions both give \(\omega > 0\) for each wave number \(q\) used. Like the results for advancing fronts, level-set predictions better match theory for long-wavelength perturbations. We further demonstrate the instability of receding fronts by computing numerical solutions beyond the initial linear regime. Results are presented in Figures~\cref{fig:results_lsa_receding}E--F. As Figure~\cref{fig:results_lsa_receding}F shows, the initial small-amplitude perturbation grows as the front recedes. Again, the discrepancy between theoretical and numerical growth rate might occur because the grid spacing \(\Delta x = \Delta y = 0.025\) cannot fully resolve small-wavelength perturbations. Indeed, the interface shape (red line in Figure~\cref{fig:results_lsa_receding}F) indicates difficulties resolving curvature in receding fronts with growing perturbations.
%
%%%%%%%%%%%%%%%%%%%%%%%%%%%%%%%%%%%%%%%%%%%%%%%
%%%%% Results: Regularised Receding Waves %%%%%
%%%%%%%%%%%%%%%%%%%%%%%%%%%%%%%%%%%%%%%%%%%%%%%
%
\subsection{Surface Tension Regularisation Stabilises Receding Waves}\label{ssec:results_receding_regularised}
Regularisations are commonly applied to obtain variations to moving-boundary models. These variations might prevent blow-up or suppress unphysical solutions. Regularising moving-boundary models can also stabilise solutions that are unstable without regularistion~\cite{Chadam1983,Kitsunezaki1997}. For example, Chadam and Ortoleva~\cite{Chadam1983} showed that planar melting in the classical Stefan problem is linearly stable to transverse perturbations of all wave numbers. Conversely, planar solidification is unstable, but becomes stable for some wave numbers if the classical Stefan problem is modified to include surface tension~\cite{Chadam1983}. Introducing surface tension involves replacing the usual interface density condition \(u(S(t),t) = 0\) with \(u(S(t),t) = \gamma K\)~\cite{Chadam1983,Back2014,Kitsunezaki1997}, where \(\gamma\) is the surface tension coefficient, and \(K\) is local curvature. We apply the surface tension regularisation developed for the classical Stefan problem to the Fisher--Stefan model. Planar melting in the classical Stefan problem is analogous to advancing (\(\kappa > 0\)) solutions in the Fisher--Stefan model, whereas planar solidification corresponds to receding solutions (\(\kappa < 0\)). In this work, we investigate whether surface tension can also stabilise the unstable receding (\(\kappa < 0\)) Fisher--Stefan fronts seen in Section~\cref{ssec:results_receding}.

Surface tension regularisation affects the density at the interface between the active and inactive populations. Previously, this was set to \(u = u_{\mathrm{f}}\) in the 2D Fisher--Stefan model~\cref{eq:model_fs_2d}. For receding solutions with \(\kappa < 0,\) we consider the regularised 2D Fisher--Stefan model,
\begin{subequations}
  \label{eq:model_fs_2d_regularised}%
  \begin{gather}
    \pd{u}{t} = \pdn{u}{x}{2} + \pdn{u}{y}{2} + u\left(1-u\right), \quad\text{ on }\quad x < L(y,t),\label{eq:model_fs_2d_regularised_fkpp}\\
    u = 1, \quad\text{ on }\quad x = 0\label{eq:model_fs_2d_regularised_bc_left},\\
    u = u_{\mathrm{f}} - \gamma K, \quad\text{ on }\quad x = L(y,t)\label{eq:model_fs_2d_regularised_bc},\\
    V = -\kappa\nabla u\cdot\mathvec{\hat{n}}, \quad\text{ on }\quad x = L(y,t)\label{eq:model_fs_2d_regularised_stefan},\\
    u(x,y,0) = U(x,y), \quad\text{ on }\quad x < L(y,0)\label{eq:model_fs_2d_regularised_ic}.
  \end{gather}
\end{subequations}
The difference between~\cref{eq:model_fs_2d_regularised,eq:model_fs_2d} is that the interface density~\cref{eq:model_fs_2d_regularised_bc} now depends on the local curvature, \(K,\) and surface tension coefficient, \(\gamma.\) The equivalent level-set form of~\cref{eq:model_fs_2d_regularised} is
\begin{subequations}
  \label{eq:model_fs_2d_level-set_analysis_regularised}%
  \begin{gather}
    \pd{u}{t} = \pdn{u}{x}{2} + \pdn{u}{y}{2} + u\left(1-u\right) \quad\text{ on }\quad \phi(x,y,t) < 0,\label{eq:model_fs_2d_level-set_analysis_regularised_fkpp}\\
    \pd{\phi}{t} = \kappa\nabla u\cdot\nabla\phi \quad\text{ on }\quad \phi(x,y,t) = 0,\label{eq:model_fs_2d_level-set_analysis_regularised_ls}\\
    u = 1, \quad\text{ on }\quad x = 0\label{eq:model_fs_2d_level-set_analysis_regularised_bc_left},\\
    u = u_{\mathrm{f}} - \gamma K \quad\text{ on }\quad \phi(x,y,t) = 0\label{eq:model_fs_2d_level-set_analysis_regularised_bc},\\
    u(x,y,0) = U(x,y) \quad\text{ on }\quad \phi(x,y,0) < 0.\label{eq:model_fs_2d_level-set_analysis_regularised_ic},
  \end{gather}
\end{subequations}
where the local (signed) curvature is
\begin{equation}
  \label{eq:results_curvature}%
  K = \nabla\cdot\frac{\nabla\phi}{\left\lvert\nabla\phi\right\rvert} = \frac{\phi_{xx}\phi_{y}^{2} - 2\phi_{y}\phi_{x}\phi_{xy} + \phi_{yy}\phi_{x}^{2}}{\left(\phi_{x}^{2} + \phi_{y}^{2}\right)^{3/2}}.
\end{equation}
In the curvature term~\cref{eq:results_curvature}, we use subscripts to denote partial differentiation for compactness.  Biologically, surface tension might represent cell--cell adhesion between cells located on the interface~\cite{Forgacs1998}. 

\begin{figure}[htbp!]
  \centering
  \includegraphics[width=\linewidth]{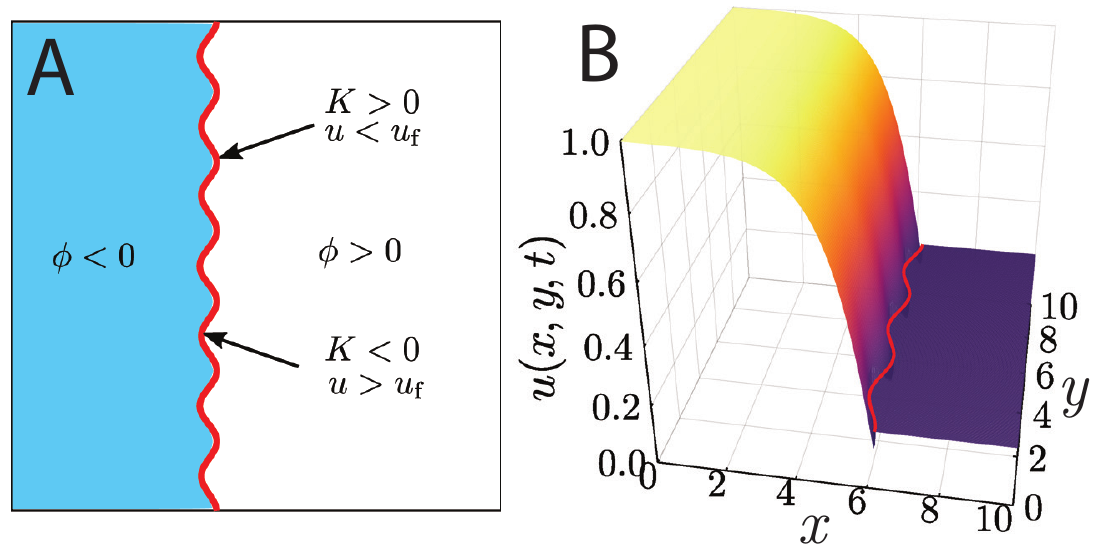}
  \caption{The linear stability ansatz with surface tension regularisation. (A) Schematic of the perturbed front illustrating regions of positive and negative curvature. Red curve indicates the zero level set (interface between occupied and vacant or partially-vacant regions). (B) Example density profile for a perturbed front with surface tension regularisation. Red curve indicates the zero level set. Parameters: \(\varepsilon = 0.2,\) \(\kappa = -0.5,\) \(\gamma = 0.1,\) \(u_{\mathrm{f}} = 0.2,\) \(\beta = 6,\) \(q = 3\pi/5.\)}
  \label{fig:results_ansatz_regularisation}
\end{figure}

The effect of introducing surface tension is illustrated in Figure~\cref{fig:results_ansatz_regularisation}. A sketch of how surface tension affects interface density is shown in Figure~\cref{fig:results_ansatz_regularisation}A. Curvature~\cref{eq:results_curvature} is defined such that \(K > 0\) when the perturbed interface is concave. This occurs when the perturbed interface is ahead (to the right) of the planar front, as Figure~\cref{fig:results_ansatz_regularisation}A shows. According to~\cref{eq:model_fs_2d_level-set_analysis_regularised_bc}, in this  \(K > 0\) scenario the population density at the interface will decrease, such that \(u < u_{\mathrm{f}}\) on \(\phi = 0.\) Conversely, \(K < 0\) when the perturbation interface is convex. As Figure~\cref{fig:results_ansatz_regularisation}A shows, this occurs when the perturbed interface is behind (to the left) of the unperturbed front. When \(K < 0,\) the interface density increases, such that \(u > u_{\mathrm{f}}\) on \(\phi = 0.\) An example density profile for a perturbed front with surface tension is shown in Figure~\cref{fig:results_ansatz_regularisation}B.
In biological contexts, population density \(u(x,y,t)\) must be non-negative everywhere. Since sinusoidal perturbations give regions where \(K < 0,\) a sufficiently large choice of \(u_{\mathrm{f}}\) is required to ensure \(u_{\mathrm{f}} - \gamma K \geq 0\) everywhere. This consideration does not apply to the classical Stefan problem with surface tension, for which the scaled temperature can be negative. Biological interpretation of surface tension thus requires care.

The surface tension regularisation modifies the linear stability analysis for \(\kappa < 0.\) Since derivatives of \(\phi\) with respect to \(y\) only appear at \(\mathcal{O}(\varepsilon)\) for perturbed planar fronts, the leading-order problem~\cref{eq:results_lo} is unchanged. The first-order correction problem becomes
\begin{subequations}
  \label{eq:results_regularised_fo}%
  \begin{gather}
    \fdn{u_1}{\xi}{2} + c\fd{u_1}{\xi} + \left[1-\omega - q^2 - 2u_0(\xi)\right]u_1(\xi) = \left(\omega + q^2\right)\fd{u_0}{\xi} \quad\text{ on }\quad \xi < 0,\label{eq:results_regularised_fo_fkpp}\\
    u_1(-\infty) = 0, \quad u_1(0) = -\gamma q^2,\label{eq:results_regularised_fo_bc}\\
    \fd{u_1(0)}{\xi} = \frac{\omega}{\kappa},\label{eq:results_regularised_fo_stefan}
  \end{gather}
\end{subequations}
where surface tension now appears in the boundary conditions~\cref{eq:results_regularised_fo_bc}. Full details are available in~\cref{app:lsa}. We solve~\cref{eq:results_regularised_fo} using the numerical shooting method described in~\cref{app:numerics_shooting}. 

\begin{figure}[htbp!]
  \centering
  \includegraphics[width=\linewidth]{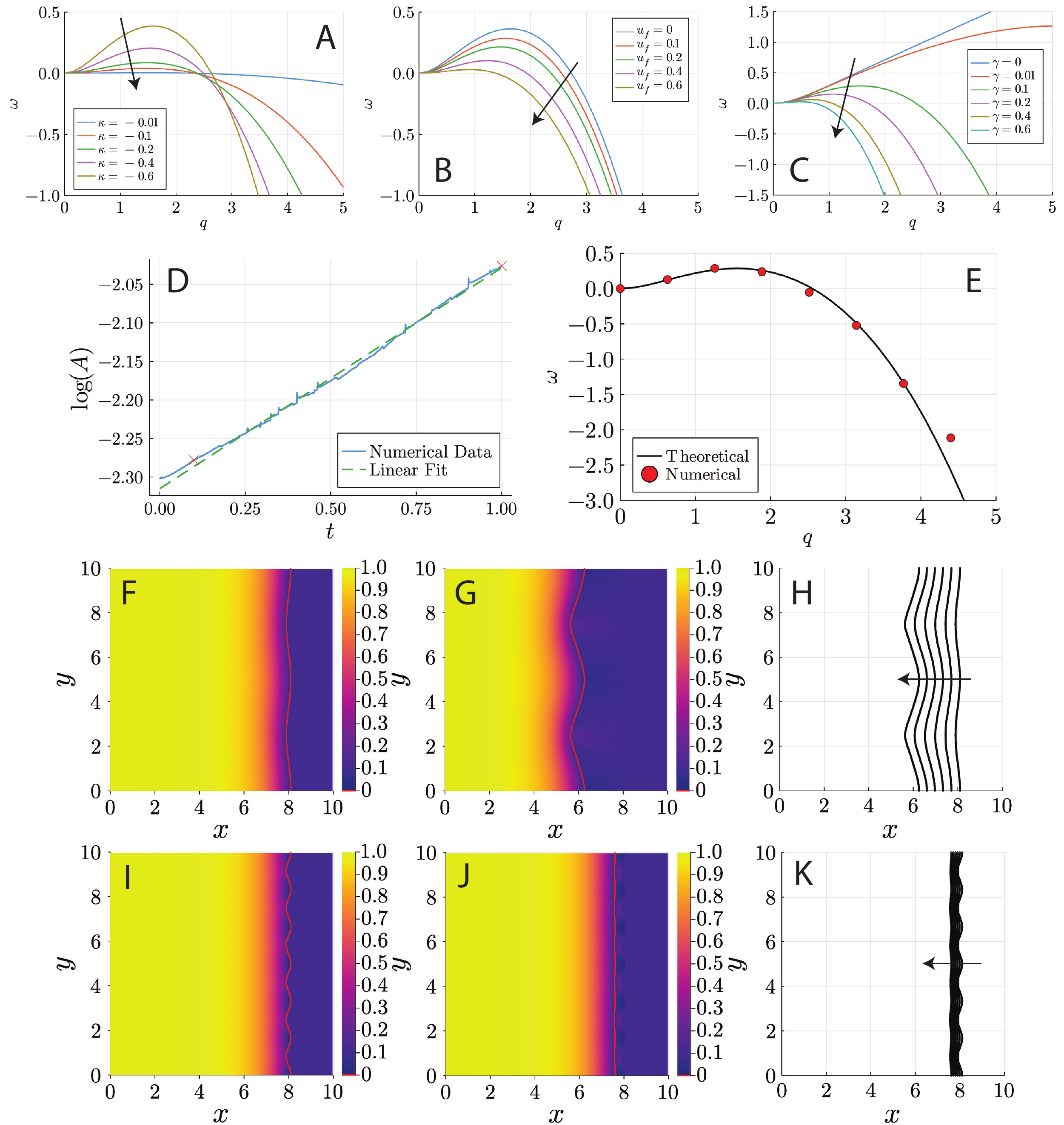}
  \caption{Linear stability analysis results for receding fronts (\(\kappa < 0\)) with the surface tension regularisation. (A--C): Dispersion relations shows the relationship between wave number, \(q,\) and growth rate \(\omega.\) For each positive wave number \(q > 0,\) the growth rate \(\omega < 0,\) indicating that the front is linearly stable to transverse perturbations. Arrows indicate direction of increasing parameter. (A) Dispersion curves with varying \(\kappa,\) \(u_{\mathrm{f}} = 0.1,\) and \(\gamma = 0.1.\) (B) Dispersion curves with \(\kappa = -0.5,\) varying \(u_{\mathrm{f}},\) and \(\gamma = 0.1.\) (C) Dispersion curves with \(\kappa = -0.5,\) \(u_{\mathrm{f}} = 0.1,\) and varying \(\gamma.\) (D--E): Comparison between theoretical and numerical dispersion relation. (D) Linear regression to determine estimate the growth rate for \(\kappa = -0.5,\) \(u_{\mathrm{f}} = 0.1,\) \(\gamma = 0.1,\) and \(q = 2\pi/5.\) Data between the two crosses are used for the linear fit. (E) Dispersion relation for \(\kappa = -0.5,\)  \(u_{\mathrm{f}} = 0.1,\) and \(\gamma = 0.1.\) Dots indicate growth rate estimated from 2D level-set numerical solutions, and the solid curve is the theoretical result. (F--K): Numerical solution to the Fisher--Stefan model with \(\kappa = -0.5,\) \(u_{\mathrm{f}} = 0.1,\) and \(\gamma = 0.1.\) Red curves denote the zero level set (interface between occupied and vacant or partially-vacant regions). (F) Initial condition~\cref{eq:results_numerical_ic}, with \(\varepsilon = 0.1,\) \(q = 2\pi/5,\) and \(\beta = 8.\) (G) Numerical solution \(u(x,y,5).\) (H) Interface plotted for \(t \in \{0,1,2,3,4,5\},\) showing front instability. Arrow indicates increasing time. (I) Initial condition~\cref{eq:results_numerical_ic}, with \(\varepsilon = 0.1,\) \(q = 6\pi/5,\) and \(\beta = 8.\) (J) Numerical solution \(u(x,y,1).\) (K) Interface plotted for \(t \in \{0,0.2,0.4,0.6,0.8,1\},\) showing front stability. Arrow indicates increasing time.}
  \label{fig:results_lsa_receding_regularised}
\end{figure}

Linear stability analysis results for the Fisher--Stefan model with surface tension are presented in Figure~\cref{fig:results_lsa_receding_regularised}(A--C). Firstly, in Figure~\cref{fig:results_lsa_receding_regularised}A, we set \(u_{\mathrm{f}} = 0.1,\) \(\gamma = 0.1,\) and vary \(\kappa.\) In Figure~\cref{fig:results_lsa_receding_regularised}A, we set \(\kappa = -0.5,\) \(\gamma = 0.1,\) and vary \(u_{\mathrm{f}}.\) Finally, in Figure~\cref{fig:results_lsa_receding_regularised}A, we set \(\kappa = -0.5,\) \(u_{\mathrm{f}} = 0.1,\) and vary \(\gamma.\) These dispersion relations illustrate the stabilising effect of surface tension on receding fronts. Unlike the unregularised results in Section~\cref{ssec:results_receding} where receding fronts were unstable for all \(q,\) with surface tension there is only a finite range of unstable wave numbers \(q \in (0, q^*).\) These unstable wave numbers correspond to long-wavelength instabilities. As wave number increases, the perturbation wavelength decreases. This increases local curvature on the interface, which sufficiently amplifies surface tension effects, stabilising the front. With surface tension included, the theory predicts existence of a most-unstable wave number, \(q_{\mathrm{max}},\) that maximises \(\omega.\) A receding front perturbed at random with multiple modes will generate a pattern with wavelength corresponding to \(q_{\mathrm{max}}.\) The most unstable wave number depends on \(\kappa,\) \(u_{\mathrm{f}},\) and \(\gamma,\) giving rise to richer dynamics than unregularised fronts.

Numerical dispersion relations predicted using level-set solutions to the 2D Fisher--Stefan model with surface tension agree well with theoretical predictions. An example with \(\kappa = -0.5,\) \(u_{\mathrm{f}} = 0.1,\) and \(\gamma = 0.1\) is shown in Figure~\cref{fig:results_lsa_receding_regularised}E. For these parameters, numerical solutions successfully predict the most unstable wave number, and the transition from instability to stability. Full numerical solutions beyond the linear regime also corroborate the analysis. Figures~\cref{fig:results_lsa_receding_regularised}F--H show a solution with \(q = 2\pi/5,\) which theory predicts to be unstable. As Figures~\cref{fig:results_lsa_receding_regularised}G--H illustrates, the perturbation amplitude grows with time. However, perturbations decay in the solution in Figures~\cref{fig:results_lsa_receding_regularised}I--K with \(q = 6\pi/5,\) for which linear theory predicts stability.
%
%%%%%%%%%%%%%%%%%%%%%%%%%%%%%%%%%%%%%
%%%%% DISCUSSION AND CONCLUSION %%%%%
%%%%%%%%%%%%%%%%%%%%%%%%%%%%%%%%%%%%%
%
\section{Discussion and Conclusion}\label{sec:discussion}
We investigated the linear stability of planar travelling wave solutions to the Fisher--Stefan model to transverse perturbations. Depending on the sign of the parameter \(\kappa,\) these travelling wave solutions can have wave speeds \(c \in (-\infty,\infty),\) whilst retaining non-negative population density. Therefore, Fisher--Stefan waves can model both population invasion and recession. Linear instability indicates the possibility of spontaneous pattern formation. We showed that advancing solutions to the Fisher--Stefan model with \(c > 0\) are linearly stable to perturbations of all wave numbers. However, receding waves with \(c < 0\) in the Fisher--Stefan model are linearly unstable for all wave numbers. For receding waves, growth rate increases as the perturbation wave length decreases, making it impossible to identify a most unstable wave number. However, regularising the receding fronts by introducing surface tension stabilises solutions with short wavelength perturbations. Regularised receding fronts then have a range of unstable wave numbers \(q \in (0, q^*).\) Furthermore, the maximum value of \(\omega(q)\) enables us to identify the most unstable wave number for regularised receding fronts. This most unstable wave number corresponds to the wavelength of the pattern expected to dominate in practice, where populations are continuously subjected to random perturbations. The Fisher--Stefan model thus provides a possible explanation for pattern formation in receding populations, and enables us to predict the characteristic length of patterns.

Throughout the work, we support and extend theoretical analysis with 2D numerical solutions to the Fisher--Stefan model. Numerical solutions corroborated the linear stability results, correctly identifying the range of stable and unstable wave numbers for both advancing and receding waves. Furthermore, our numerical results showed that analytical predictions of instability and stability remain valid beyond the initial linear regime. Our numerical solutions were obtained using the level-set method, as described in~\cref{app:numerics_level-set} and online in our \href{https://github.com/alex-tam/2D_Fisher-Stefan_Level-Set_Stability}{GitHub repository}. When used to predict the growth rate in numerical solutions, our numerical results exhibit some discrepancy from the theoretical results, particularly for unregularised receding fronts and for larger wave numbers. These discrepancies might occur due to insufficient grid resolution, or because the numerical solutions extend beyond the short time and small-amplitude perturbation regime for which linear theory applies. However, our numerical method successfully captures both stability and instability of regularised fronts, and accurately reproduces the dispersion relation in regularised receding fronts.

A key feature of the Fisher--Stefan model is that it admits receding travelling wave solutions with \(c < 0\) and non-negative population density. In contrast, all biologically-relevant travelling wave solutions of the standard Fisher--KPP model advance, with \(c \geq 2.\) The Fisher--Stefan model and other reaction--diffusion moving-boundary problems offer a new approach for modelling populations that recede. For example, a receding travelling wave in the regularised Fisher--Stefan model might represent a shrinking tumour. Cell--cell adhesion on the tumour boundary might then be analogous to surface tension. A planar wave approximate the shape of a large tumour with low boundary curvature~\cite{Tam2018}. The most unstable wave number then predicts characteristic shape of a tumour as it recedes. Under this interpretation, linear instability of receding waves enables predictions of biological pattern formation that were not possible in the Fisher--KPP model.

Our work also unites theory for the Fisher--KPP equation and classical Stefan problem. Planar fronts in the Fisher--KPP equation are linearly stable~\cite{Huang2008,Zeng2014}. These fronts are similar to advancing waves \(\kappa, c > 0\) in the Fisher--Stefan model, which we also found to be stable. Receding fronts in the Fisher--Stefan model are analogous to planar solidification in the classical Stefan problem. Since solidification fronts are unstable~\cite{Chadam1983}, it is unsurprising that receding Fisher--Stefan fronts are also unstable. Although the stability analysis results in the Fisher--Stefan model are similar to planar solidification in the classical Stefan problem, interpreting the analysis requires great care. In the classical Stefan problem, the temperature \(u(x,t)\) is scaled such that \(u_{\mathrm{f}} = 0\) is the melting temperature, and \(u = 1\) is the far-field temperature. After applying a surface tension regularisation similar to~\cref{eq:model_fs_2d_regularised_bc}, the temperature will be negative (that is, below melting temperature) wherever \(K > 0.\) In contrast, \(u(\mathvec{x},t)\) represents a biological population in the Fisher--Stefan model. Such a population cannot be negative. To circumvent this potential difficulty, we allow biological populations to have \(u_{\mathrm{f}} > 0.\) Applying~\cref{eq:model_fs_2d_regularised_bc} then does not introduce negative population density for fronts with sufficiently low curvature. This non-zero \(u_{\mathrm{f}}\) at the interface might represent a partially-cleared wound, or cell population invading into a non-vacant region.

This research opens avenues for future work. One extension would be to apply the stability analysis methods to other reaction--diffusion--Stefan problems with general nonlinear reaction and diffusion terms~\cite{Fadai2021}. Nonlinear reaction--diffusion equations have sharp-fronted travelling wave solutions that advance~\cite{Muller2002,Tam2018}. Unlike the linear diffusion Fisher--KPP equation, planar fronts with nonlinear diffusion can be unstable~\cite{Muller2002}. Applying linear stability analysis would enable us to determine whether advancing fronts with nonlinear diffusion remain unstable in the moving-boundary framework. Linear stability analysis could also be used to understand two-dimensional patterns in other moving-boundary problems, including multiple-species models~\cite{El-Hachem2020}. Another extension would be to consider pattern formation in non-planar geometry. The instabilities found for planar receding waves might also apply to general hole-closing geometries. For example, investigating closure of a circular hole in polar co-ordinates might better approximate the geometry of a shrinking wound or tumour. In addition, it is unknown whether one-dimensional travelling wave solutions to the Fisher--Stefan model are themselves stable. This is, if we perturb the travelling wave density profile, will the one-dimensional wave persist? We plan to address these questions in future studies.
%
%%%%%%%%%%%%%%%%%%%%%%%
%%%%% Back Matter %%%%%
%%%%%%%%%%%%%%%%%%%%%%%
%
\section*{CRediT Authorship Contribution Statement}
{\bfseries Alexander K. Y. Tam:} Designed the research. Wrote the numerical code. Performed the analysis. Obtained the results. Wrote the manuscript. {\bfseries Matthew J. Simpson:} Designed the research, edited the manuscript.
\section*{Declaration of Competing Interests}
The authors declare that they have no known competing financial interests or personal relationships that could have appeared to influence the work reported in this paper.
\section*{Acknowledgements}
M. J. S. acknowledges funding from the Australian Research Council (Grant No. DP200100177).
%
%%%%%%%%%%%%%%%%%%%%%%%%
%%%%% BIBLIOGRAPHY %%%%%
%%%%%%%%%%%%%%%%%%%%%%%%
%
\bibliography{Fisher-Stefan_Stability}
%
%%%%%%%%%%%%%%%%%%%%
%%%%% APPENDIX %%%%%
%%%%%%%%%%%%%%%%%%%%
%
\appendix
%
%%%%%%%%%%%%%%%%%%%%%%%%%%%%%%%%%%%%%%%%%%%%%%%
%%%%% Appendix: Linear Stability Analysis %%%%%
%%%%%%%%%%%%%%%%%%%%%%%%%%%%%%%%%%%%%%%%%%%%%%%
%
\clearpage
\section{Linear Stability Analysis}\label{app:lsa}
This Appendix contains full details of the linear stability analysis performed in Section~\cref{sec:results}. The analysis starts from the level-set formulation of the 2D Fisher--Stefan model. With surface tension regularisation included, this is
\begin{subequations}
  \label{eq:app_fs_2d}%
  \begin{gather}
    \pd{u}{t} = \pdn{u}{x}{2} + \pdn{u}{y}{2} + u\left(1-u\right) \quad\text{ on }\quad \phi(x,y,t) < 0,\label{eq:app_fs_2d_fkpp}\\
    \pd{\phi}{t} = \kappa\left(\pd{u}{x}\pd{\phi}{x} + \pd{u}{y}\pd{\phi}{y}\right) \quad\text{ on }\quad \phi(x,y,t) = 0,\label{eq:app_fs_2d_ls}\\
    u = 1 \quad\text{ on }\quad x = 0,\label{eq:app_fs_2d_bc_left}\\
    u = u_{\mathrm{f}} - \gamma\left[\frac{\phi_{xx}\phi_{y}^{2} - 2\phi_{y}\phi_{x}\phi_{xy} + \phi_{yy}\phi_{x}^{2}}{\left(\phi_{x}^{2} + \phi_{y}^{2}\right)^{3/2}}\right]\quad\text{ on }\quad \phi(x,y,t) = 0,\label{eq:app_fs_2d_bc}\\
    u(x,y,0) = U(x,y) \quad\text{ on }\quad \phi(x,y,0) < 0,\label{eq:app_fs_2d_ic}
  \end{gather}
\end{subequations}
where subscripts in~\cref{eq:app_fs_2d_bc} denote partial differentiation. To implement the linear stability analysis, we introduce the change of variables
\begin{equation}
  \label{eq:app_lsa_cov}%
  \xi = x - ct - \varepsilon\e^{\i qy + \omega t}, \quad \eta = y, \quad \tau = t,
\end{equation}
where \(\varepsilon \ll 1.\) The variable \(\xi\) follows a planar travelling wave solution moving rightward with speed \(c,\) that is perturbed with a small-amplitude sinusoidal perturbation of wave number \(q.\) Furthermore, we assume that \(\xi = 0\) corresponds to the interface position, such that \(\phi = 0.\) Therefore, to facilitate the analysis we also set
\begin{equation}
  \label{eq:app_lsa_phi}%
  \phi =  x - ct - \varepsilon\e^{\i qy + \omega t} + \mathcal{O}(\varepsilon^2).
\end{equation}
The change of variables~\cref{eq:app_lsa_cov} then requires
\begin{subequations}
  \label{eq:app_lsa_cov_derivatives}%
    \begin{align}
        \pd{u}{x} &= \pd{u}{\xi},\label{eq:app_lsa_cov_derivatives_x}\\
        \pdn{u}{x}{2} &= \pdn{u}{\xi}{2},\label{eq:app_lsa_cov_derivatives_xx}\\
        \pd{u}{y} &= \pd{u}{\eta} - \varepsilon\i q\e^{\i q\eta + \omega\tau}\pd{u}{\xi},\label{eq:app_lsa_cov_derivatives_y}\\
        \pdn{u}{y}{2} &= \pdn{u}{\eta}{2} + 2\varepsilon\i q\e^{\i q\eta+\omega\tau}\pdt{u}{\xi}{\eta} + \varepsilon q^2\e^{\i q\eta+\omega\tau}\pd{u}{\xi} ,\label{eq:app_lsa_cov_derivatives_yy}\\
        \pd{u}{t} &=\pd{u}{\tau} - c\pd{u}{\xi} - \varepsilon\omega\e^{\i q\eta + \omega\tau}\pd{u}{\xi}.\label{eq:app_lsa_cov_derivatives_t}
    \end{align}
\end{subequations}
In terms of the new variables \((\xi,\eta,\tau),\) we expand the population density as
\begin{equation}
  \label{eq:app_lsa_expansion}%
  u(\xi, \eta,\tau) = u_0(\xi) + \varepsilon u_1(\xi)\e^{\i q\eta + \omega\tau} + \mathcal{O}(\varepsilon^2).
\end{equation}
Using the ansatz~\cref{eq:app_lsa_expansion}, we have
\begin{subequations}
  \label{eq:app_lsa_cov_derivatives_transformed}%
    \begin{align}
      \pd{u}{\tau} &= \varepsilon\omega u_1(\xi)\e^{\i q\eta + \omega\tau} + \mathcal{O}(\varepsilon^2),\label{eq:app_lsa_cov_derivatives_transformed_tau}\\
      \pd{u}{\xi} &= \pd{u_0}{\xi} + \varepsilon\pd{u_1}{\xi}\e^{\i q\eta + \omega\tau} + \mathcal{O}(\varepsilon^2),\label{eq:app_lsa_cov_derivatives_transformed_xi}\\
      \pdn{u}{\xi}{2} &= \pdn{u_0}{\xi}{2} + \varepsilon\pdn{u_1}{\xi}{2}\e^{\i q\eta + \omega\tau} + \mathcal{O}(\varepsilon^2),\label{eq:app_lsa_cov_derivatives_transformed_xi_xi}\\
      \pd{u}{\eta} &= \varepsilon\i qu_1(\xi)\e^{\i q\eta + \omega\tau} + \mathcal{O}(\varepsilon^2),\label{eq:app_lsa_cov_derivatives_transformed_eta}\\
      \pdn{u}{\eta}{2} &= -\varepsilon q^2u_1(\xi)\e^{\i q\eta + \omega\tau} + \mathcal{O}(\varepsilon^2),\label{eq:app_lsa_cov_derivatives_transformed_eta_eta}\\
      \pdt{u}{\xi}{\eta} &= \varepsilon\i q\pd{u_1}{\xi}\e^{\i q\eta + \omega\tau} + \mathcal{O}(\varepsilon^2).\label{eq:app_lsa_cov_derivatives_transformed_xi_eta} 
    \end{align}
\end{subequations}
Using~\cref{eq:app_lsa_cov,eq:app_lsa_phi,eq:app_lsa_cov_derivatives,eq:app_lsa_cov_derivatives_transformed}, we can rewrite the 2D Fisher--Stefan model~\cref{eq:app_fs_2d} as
\begin{subequations}
  \label{eq:app_lsa_fs_transformed}%
  \begin{gather}
      \begin{gathered}
        \varepsilon\omega u_1\e^{\i q\eta+\omega\tau} - c\fd{u_0}{\xi} - \varepsilon c\pd{u_1}{\xi}\e^{\i q\eta + \omega\tau}- \varepsilon\omega\e^{\i q\eta + \omega\tau}\fd{u_0}{\xi} \\= \fdn{u_0}{\xi}{2} + \varepsilon\fdn{u_1}{\xi}{2}\e^{\i q\eta+\omega\tau} - \varepsilon q^2u_1\e^{\i q\eta + \omega\tau} + \varepsilon q^2\e^{\i q\eta + \omega\tau}\fd{u_0}{\xi} \\+ u_0\left(1-u_0\right) + \varepsilon\left(u_1\e^{\i q\eta+\omega\tau} - 2u_0u_1\e^{\i q\eta+\omega\tau}\right) \quad \text{ on } \quad \xi < 0,
      \end{gathered}\label{eq:app_lsa_fs_transformed_fkpp}\\
      -c -\varepsilon\omega\e^{\i q\eta+\omega\tau} = \kappa\left(\fd{u_0}{\xi} + \varepsilon\fd{u_1}{\xi}\e^{\i q\eta + \omega\tau} \right) \quad \text{ on } \quad \xi = 0,\label{eq:app_lsa_fs_transformed_ls}\\
      u_0 + \varepsilon u_1\e^{\i q\eta + \omega\tau} = 1 \quad \text{ on } \quad \xi \to -\infty,\label{eq:app_lsa_fs_transformed_bc_left}\\
      u_0 + \varepsilon u_1\e^{\i q\eta+\omega\tau} = u_{\mathrm{f}} - \varepsilon\gamma\left(q^2\e^{\i q\eta + \omega\tau}\right) \quad \text{ on } \quad \xi = 0,\label{eq:app_lsa_fs_transformed_bc}
  \end{gather}
\end{subequations}
up to \(\mathcal{O}(\varepsilon)\) as \(\varepsilon \to 0.\) The leading-order problem is subsequently
\begin{subequations}
    \label{eq:app_lsa_lo}%
    \begin{gather}
        \fdn{u_0}{\xi}{2} + c\fd{u_0}{\xi} + u_0\left(1-u_0\right) = 0\quad \text{ on } \quad \xi < 0,\\
        u_0(-\infty) = 1 \quad u_0(0) = u_{\mathrm{f}},\\
        \fd{u_0(0)}{\xi} = -\frac{c}{\kappa}.
    \end{gather}
\end{subequations}
The problem for the first-order corrections is
\begin{subequations}
    \label{eq:app_lsa_fo}%
    \begin{gather}
        \fdn{u_1}{\xi}{2} + c\fd{u_1}{\xi} + \left[1-\omega-q^2 - 2u_0\right]u_1 = -\left(\omega+q^2\right)\fd{u_0}{\xi} \quad \text{ on } \quad \xi < 0,\\
        u_1(-\infty) = 0 \quad u_1(0) = -\gamma q^2,\\
        \fd{u_1}{\xi} = -\frac{\omega}{\kappa}\quad \text{ on } \quad \xi = 0.
    \end{gather}
\end{subequations}
These complete the derivation of~\cref{eq:results_lo,eq:results_fo} for the linear stability analysis.
%
%%%%%%%%%%%%%%%%%%%%%%%%%%%%%%%%%%%%%%%%%%%%%
%%%%% Appendix: Numerical Methods (BVP) %%%%%
%%%%%%%%%%%%%%%%%%%%%%%%%%%%%%%%%%%%%%%%%%%%%
%
\clearpage
\section{Numerical Methods for Boundary-Value Problems}\label{app:numerics_shooting}
This Appendix contains details of the numerical methods used for the boundary-value problems arising from the linear stability analysis. We first describe the method to solve the boundary-value problem~\cref{eq:results_lo}. First, we truncate the semi-infinite domain \(-\infty < \xi < 0\) to the finite domain \(-\xi_{\mathrm{max}} < \xi < 0.\)  The problem to solve numerically is then (dropping subscripts on the leading-order population density)
\begin{subequations}
  \label{eq:app_shooting_bvp_lo}%
  \begin{gather}
    \fdn{u}{\xi}{2} + c\fd{u}{\xi} + u\left(1-u\right) = 0 \quad\text{ on } \quad -\xi_{\mathrm{max}} < \xi < 0,\label{eq:app_shooting_bvp_lo_fkpp}\\
    u(-20) = 1, \quad u(0) = u_{\mathrm{f}},\label{eq:app_shooting_bvp_lo_bc}\\
    \fd{u(0)}{\xi} = -\frac{c}{\kappa},\label{eq:app_shooting_bvp_lo_stefan}
  \end{gather}
\end{subequations}
on \(\xi \in [-\xi_{\mathrm{max}}, 0].\) The system~\cref{eq:app_shooting_bvp_lo} is a two-point, second-order boundary value problem with three boundary conditions. Although~\cref{eq:app_shooting_bvp_lo} is overdetermined, the wave speed \(c\) is an unknown parameter. We use the shooting method to solve~\cref{eq:app_shooting_bvp_lo} and obtain the correct wave speed, \(c.\) We first solve~\cref{eq:app_shooting_bvp_lo_fkpp} subject to the two boundary conditions~\cref{eq:app_shooting_bvp_lo_bc} and an initial guess for \(c.\) Then, we use the Newton--Raphson method to obtain the unique value of \(c\) such that the third condition~\cref{eq:app_shooting_bvp_lo_stefan} holds.

To solve~\cref{eq:app_shooting_bvp_lo_fkpp,eq:app_shooting_bvp_lo_bc}, we introduce the equispaced discrete grid \(\xi_i = -\xi_{\mathrm{max}} + i\Delta\xi,\) for \(i = 0, \dots, N_\xi,\) where \(\Delta\xi = \xi_{\mathrm{max}}/N_\xi\) is constant. We discretise~\cref{eq:app_shooting_bvp_lo_fkpp} using the second-order central difference scheme
\begin{subequations}
  \label{eq:app_shooting_bvp_lo_fkpp_discretisation}%
  \begin{gather}
    u_0 = 1,\label{eq:app_shooting_bvp_lo_fkpp_discretisation_left}\\
    \frac{u_{i+1} - 2u_{i} + u_{i-1}}{(\Delta\xi)^2} + c\frac{u_{i+1} - u_{i-1}}{2\Delta\xi} + u_{i}\left(1-u_{i}\right) = 0, \quad i = 1, \dots, N_{\xi}-1,\label{eq:app_shooting_bvp_lo_fkpp_discretisation_interior}\\
    u_{N_{\xi}} = u_{\mathrm{f}}.\label{eq:app_shooting_bvp_lo_fkpp_discretisation_right}
  \end{gather}
\end{subequations}
Equations~\cref{eq:app_shooting_bvp_lo_fkpp_discretisation} forms a system of \(N_{\xi}+1\) equations for the \(N_{\xi}+1\) unknowns \(u_i.\) We solve~\cref{eq:app_shooting_bvp_lo_fkpp_discretisation} using the Newton--Raphson method. This enables us to compute the leading-order density \(u_0\) for a given \(c.\)

We use the shooting method to obtain the wave speed \(c\) that satisfies~\cref{eq:app_shooting_bvp_lo_stefan}. First, we make an initial guess \(c^{(0)},\) and solve~\cref{eq:app_shooting_bvp_lo_fkpp_discretisation}. We then introduce the residual function
\begin{equation}
  \label{eq:app_shooting_bvp_lo_residual}%
  f(c) = \frac{3u_{N_{\xi}}-4u_{N_{\xi-1}}+u_{N_{\xi-2}}}{2\Delta\xi} + \frac{c}{\kappa},
\end{equation}
such that the root \(f(c) = 0\) corresponds to the value of \(c\) satisfying~\cref{eq:app_shooting_bvp_lo_stefan}. The first term on the right-hand side of~\cref{eq:app_shooting_bvp_lo_residual} is a second-order one-sided difference formula for \(\d u/\d\xi,\) where \(u_i\) is obtained by solving~\cref{eq:app_shooting_bvp_lo_fkpp_discretisation} with the given value of \(c.\) Starting from the initial guess \(c^{(0)},\) we apply Newton--Raphson iteration
\begin{equation}
  \label{eq:app_shooting_bvp_lo_newton}%
  c^{(k+1)} = c^{(k)} - \frac{f\left(c^{(k)}\right)}{f'\left(c^{(k)}\right)}
\end{equation}
to solve \(f(c) = 0\) numerically. We approximate the derivative of \(f\) using
\begin{equation}
  \label{eq:app_shooting_bvp_lo_derivative}%
  f'(c) = \frac{f(c+\epsilon) - f(c)}{\epsilon}
\end{equation}
for \(\epsilon \ll 1,\) and terminate the Newton--Raphson iteration when \(|c^{(k+1)}-c^{(k)}| < 1 \times 10^{-6}.\) After terminating, we accept \(c^{(k+1)}\) as the correct wave speed \(c,\) and solve~\cref{eq:app_shooting_bvp_lo_fkpp_discretisation} to obtain the leading-order density profile \(u_0.\) In all solutions, we use \(\xi_{\mathrm{max}} = 20,\) \(N_{\xi} = 1000\) and \(\epsilon = 1\times 10^{-6}.\)

The method to solve the boundary-value problem~\cref{eq:results_regularised_fo} for the first-order corrections is similar to the above method for the leading-order problem. On the same truncated domain \(\xi \in [-20,0],\) the numerical problem to solve is
\begin{subequations}
  \label{eq:app_shooting_bvp_fo}%
  \begin{gather}
    \fdn{\hat{u}}{\xi}{2} + c\fd{\hat{u}}{\xi} + \left[1-\omega - q^2 - 2u(\xi)\right]\hat{u}(\xi) = -\left(\omega + q^2\right)\fd{u}{\xi} \quad\text{ on }\quad -\xi_{\mathrm{max}}< \xi < 0,\label{eq:app_shooting_bvp_fo_fkpp}\\
    \hat{u}(-20) = 0, \quad \hat{u}(0) = -\gamma q^2,\label{eq:app_shooting_bvp_fo_bc}\\
    \fd{\hat{u}(0)}{\xi} = -\frac{\omega}{\kappa},\label{eq:app_shooting_bvp_fo_stefan}
  \end{gather}
\end{subequations}
where \(\hat{u}\) denotes the first-order corrections, and \(u\) is the known leading-order solution. The central difference scheme to solve~\cref{eq:app_shooting_bvp_fo_fkpp} subject to~\cref{eq:app_shooting_bvp_fo_bc} is then
\begin{subequations}
  \label{eq:app_shooting_bvp_fo_fkpp_discretisation}%
  \begin{gather}
    \hat{u}_0 = 0,\label{eq:app_shooting_bvp_fo_fkpp_discretisation_left}\\
    \begin{gathered}
      \frac{\hat{u}_{i+1} - 2\hat{u}_{i} + \hat{u}_{i-1}}{(\Delta\xi)^2} + c\frac{\hat{u}_{i+1} - \hat{u}_{i-1}}{2\Delta\xi} + \left(1-\omega-q^2-2u_{i}\right)\hat{u}_{i} \\= -\left(\omega + q^2\right)\frac{u_{i+1} - u_{i-1}}{2\Delta\xi}, \quad i = 1, \dots, N_{\xi}-1,
    \end{gathered}\label{eq:app_shooting_bvp_fo_fkpp_discretisation_interior}\\
    \hat{u}_{N_{\xi}} = -\gamma q^2.\label{eq:app_shooting_bvp_fo_fkpp_discretisation_right}
  \end{gather}
\end{subequations}
In~\cref{eq:app_shooting_bvp_fo_fkpp_discretisation}, \(\gamma\) is the constant surface tension coefficient, \(u_i\) and \(c\) are known from the leading-order problem, and \(q\) is a fixed wave number. We subsequently apply the Newton--Raphson method to find the value of \(\omega\) that satisfies~\cref{eq:app_shooting_bvp_fo_stefan}. We achieve this by introducing
\begin{equation}
  \label{eq:app_shooting_bvp_fo_residual}%
  g(\omega) = \frac{3\hat{u}_{N_{\xi}}-4\hat{u}_{N_{\xi-1}}+\hat{u}_{N_{\xi-2}}}{2\Delta\xi} + \frac{\omega}{\kappa}, \quad g'(\omega) = \frac{g(\omega+\epsilon) - g(\omega)}{\epsilon},
\end{equation}
and applying the iteration
\begin{equation}
  \label{eq:app_shooting_bvp_fo_newton}%
  \omega^{(k+1)} = \omega^{(k)} - \frac{g\left(\omega^{(k)}\right)}{g'\left(\omega^{(k)}\right)}
\end{equation}
to solve \(g(\omega) = 0\) numerically. This provides the growth rate, \(\omega,\) corresponding the perturbation with wave number \(q.\) Computing \(\omega(q)\) for varying \(q\) then yields the dispersion curves presented in Section~\cref{sec:results}.
%
%%%%%%%%%%%%%%%%%%%%%%%%%%%%%%%%%%%%%%%%%%%%%%%%%%
%%%%% Appendix: Numerical Method (Level-Set) %%%%%
%%%%%%%%%%%%%%%%%%%%%%%%%%%%%%%%%%%%%%%%%%%%%%%%%%
%
\clearpage
\section{Level-Set Numerical Method}\label{app:numerics_level-set}
Please see the electronic Supplementary Data for full details of our level-set numerical method.
\end{document}